\documentclass{aastex62}

\graphicspath{{./}{figures/}}

\received{September 10, 2018}
\accepted{January 7, 2019}

\shorttitle{Taurus binaries with ALMA}
\shortauthors{Akeson et al.}

\begin{document}

\title{Resolved young binary systems and their disks}

\correspondingauthor{Rachel Akeson}
\email{rla@ipac.caltech.edu, ejensen1@swarthmore.edu}

\author[0000-0001-9674-1564]{Rachel L. Akeson}
\affiliation{IPAC-NExScI, Caltech, Pasadena, CA 91125, USA}

\author[0000-0002-4625-7333]{Eric L. N. Jensen}
\affiliation{Swarthmore College, Dept.\ of Physics \& Astronomy, 500 College Ave., Swarthmore, PA 19081-1390, USA}

\author[0000-0003-2251-0602]{John Carpenter}
\affiliation{Joint ALMA Observatory, Avenida Alonso 3107 Vitacura, Santiago, Chile}

\author{Luca Ricci}
\affiliation{Department of Physics and Astronomy, California State University Northridge, 18111 Nordhoff Street, Northridge, CA 91330, USA}

\author[0000-0001-8407-2105]{Stefan Laos}
\affiliation{Physics and Astronomy Department, Vanderbilt University, 6301 Stevenson Center
Nashville, TN 37235}

\author[0000-0003-0066-546X]{Natasha F. Nogueira}
\affiliation{Swarthmore College, Dept.\ of Physics \& Astronomy, 500 College Ave., Swarthmore, PA 19081-1390, USA}

\author[0000-0003-1253-5240]{Emma M. Suen-Lewis}
\affiliation{Swarthmore College, Dept.\ of Physics \& Astronomy, 500 College Ave., Swarthmore, PA 19081-1390, USA}

\begin{abstract}
    We have conducted a survey of young single and multiple systems in the Taurus-Auriga star-forming region with the Atacama Large Millimeter Array (ALMA), substantially improving both the spatial resolution and sensitivity with which individual protoplanetary disks in these systems have been observed. These ALMA observations can resolve binary separations as small as 25--30 AU and have an average 3$\sigma$ detection level of 0.35 mJy, equivalent to a disk mass of $4 \times 10^{-5}$ M$_{\odot}$ for an M3 star. Our sample was constructed from stars that have an infrared excess and/or signs of accretion and have been classified as Class II\null.  For the binary and higher order multiple systems observed, we detect $\lambda = 1.3$ mm continuum emission from one or more stars in all of our target systems.    Combined with previous surveys of Taurus, our 21 new detections increase the fraction of millimeter-detected disks to over 75\% in all categories of stars (singles, primaries, and companions)  earlier than spectral type M6 in the Class II sample.  Given the wealth of other information available for these stars, this has allowed us to study the impact of multiplicity with a much larger sample.  While millimeter flux and disk mass are related to stellar mass as seen in previous studies, we find that both primary and secondary stars in binary systems with separations of 30 to 4200 AU have lower values of millimeter flux as a function of stellar mass than single stars. We also find that for these systems, the circumstellar disk around the primary star does not dominate the total disk mass in the system and contains on average 62\% of the total mass.
    
\end{abstract}

\keywords{protoplanetary disks, binaries: general, stars: pre-main sequence}

\section{Introduction}

The formation, evolution, and dissipation of circumstellar disks are key components in understanding the formation of stellar and planetary systems, but despite years of study, some puzzling questions about circumstellar disks remain. One of the chief questions is why stars of similar ages, in the same star-forming region, can have very different disk properties. Millimeter interferometry has been crucial in confirming the paradigm of a Keplerian rotating disk of gas and dust that funnels material onto the central star, and millimeter continuum flux is the most sensitive probe of cold dust in the outer disk \citep[see e.g.][]{wil11}. Here we present results from an ALMA survey, taking advantage of the unprecedented combination of high sensitivity and angular resolution to explore two of the factors affecting these disks: the influence of stellar mass, and the influence of stellar companions. 

Previous work has demonstrated that at a given stellar mass, 
the mass of the circumstellar disk ranges over more than an order of magnitude in Taurus \citep{and13}, but many non-detections remain at the same flux level as the lower flux detections.  At lower stellar masses in particular (later than M3), the sample is dominated by non-detections and while the non-detections are consistent with the $L_{\rm disk} \sim M^{1.5-2.0}$ fit of \citet[see also \citealt{pas16}]{and13}, our ALMA Cycle 0 observations of wide binaries in Taurus \citep[hereinafter Paper 1]{ake14} revealed several disks at flux levels below the sensitivity of pre-ALMA surveys.  

Early studies of the impact of multiplicity generally did not resolve the individual disks, but did show a decrease in flux for binaries with separations of a few to $\sim100$ AU \citep{ost95,jen96}. 
Initial interferometric observations to resolve the separate circumprimary and circumsecondary disks detected the primary disk, but only rarely detected the secondary disk and were limited to small ($<$5) sample sizes \citep{jen03,pat06}.  Only the advent of high resolution and high sensitivity millimeter surveys provided sufficient samples to detect more individual components.  In the Taurus star forming region, two recent studies have concentrated on disks in multiple systems.  \citet{har12} used the {\it Submillimeter Array} to observe 23 multiple systems in Taurus.  They found a lower detection rate for stars in multiple systems (28--37\%) as compared to single stars (62\%) and a correlation of larger binary separation with higher flux.
In Paper 1, we described an ALMA survey of 17 Class II binaries in Taurus which detected 10 secondary disks and found that within binary systems the primary/secondary stellar mass ratio is not correlated with the primary/secondary flux or disk mass ratio.

Despite the significant amount of previous work on this issue, some questions remain open.  In particular, the sensitivity level of pre-ALMA surveys was often insufficient to detect disks around lower-mass stars, given the well-known correlation between stellar mass and millimeter flux \citep[e.g.][]{and13}. As a result, these surveys were less sensitive to secondary stars in binaries (by definition of lower stellar mass than their primary counterparts), and they were often quite incomplete for the low-mass tail of the single star population as well, leading to biases when comparing single stars to secondaries.  Thus, in designing this ALMA study to probe both the influence of stellar mass and of companions on circumstellar disks, we specifically included both undetected single stars and all previously unresolved multiple systems where the component separations were resolvable in a snapshot survey at moderate (i.e.\ $\sim$0\farcs2) resolution. When combined with
previous detections from the literature and Paper 1, we have compiled a significant sample of primary and secondary components to compare to their single star counterparts. 

Our ALMA observations are described in \S 2, and the results of our survey, including calculation of the circumstellar disk mass, in \S 3.  The definition of a carefully selected sample for statistical analysis is given in \S 4, along with the comparison of disk properties between single and binary stars.  Our conclusions are given in \S 5.

\section{Observations}

\subsection{Sample}

We selected 
targets from a single star forming region, Taurus (distance $\sim$ 140 pc), so that 
effects such as age and cluster environment are kept constant as much as 
possible.   Taurus is ideal in having a significant population of YSOs that 
have evolved into the disk-only state (with no remaining envelope) and in being 
very well studied, containing both a well-known set of single stars with 
disks and a significant population of binaries where both 
stellar components have been characterized in the optical or near-infrared.  
We started with the list of Taurus objects from \citet{luh10} and
selected those with Class II spectral energy distributions (SEDs), which resulted in 211 stars in 166 systems. Twelve of these stars are spectroscopic binaries. 
Some of the multiple star systems were classified using separate SEDs, but most of the binary star systems are unresolved in the mid-infrared and were classified as a pair.  
We then eliminated all single and
close ($<0\farcs18$ or 25 AU) multiple sources that had been previously detected at
millimeter wavelengths,  as well as multiple systems where all
resolvable (i.e.\ separations $> 0\farcs18$) components have been detected 
\citep[Paper 1]{har12,and13}.  We also removed sources with 
observations in ALMA Cycle 1.  The remaining list included 69 single and multiple systems with 94 resolvable disks, including 1 G star, 7 K stars, and the rest M stars. 

\subsection{ALMA Data Reduction}

To construct the observing groups for ALMA, 
the sample was divided into close multiple systems (separations $<1\arcsec$), and singles and wide multiples, to allow for different spatial
resolution observations.  Within these groups, the sample was split to follow
the ALMA guidelines on maximum angular separation from the
gain calibrator.  These divisions resulted in five source
groupings; four of these were observed in Cycle 2, and one of the four was 
re-observed in Cycle 3.  Table \ref{tab:sets} lists the basic
information for these observation sets.  One source group was never
observed.  The data obtained included observations of 45 systems, with a total of 65 stars. 

\begin{deluxetable}{llll}
\tabletypesize{\scriptsize}
\tablewidth{0pt}
\tablecaption{Observation Log \label{tab:sets}
}
\tablehead{
\colhead{Project code}    & \colhead{Observation Date}   & \colhead{Antennas} & \colhead{Beam (arcs)} 
}
\startdata
2013.1.00105.S & 2015 May 3\tablenotemark{a}&  36 & 1.7 x 0.9 \\
2013.1.00105.S & 2015 May 3 & 36 & 0.18 x 0.16 \\
2013.1.00105.S & 2015 Sept 18 & 34 & 0.24 x 0.13  \\
2013.1.00105.S & 2015 Sept 19 & 36 & 0.22 x 0.14  \\
2015.1.00392.S & 2016 July 1\tablenotemark{b} & 41 & 0.75 x 0.42  \\
\enddata
\tablenotetext{a}{High rms; data not used.}
\tablenotetext{b}{Repeat of first 2015 May 3 dataset; used in analysis here.}
\end{deluxetable}

We selected Band 6 (1.3 mm) for these ALMA observations. Three of the correlator sections were set for continuum emission sensitivity, with the fourth set to the transition for CO(2--1) at 230.5 GHz.  The total continuum bandwidth was 7.5 GHz.  For the datasets from 2015, we used the calibration provided by ALMA and created images using the {\it CASA} package \citep{mcm07}.  For sources with sufficient continuum flux, we also performed self-calibration. The data taken in July 2016 did not pass the internal quality assessment at ALMA, so we processed the raw data using the pipeline scripts in {\it CASA}. Based on the gain calibration and measured fluxes for sources with previous measurements, we deemed the data usable and they are included in the analysis below.

For each target, the {\it CASA} routine {\it clean} was used to produce an image.  As the source positions are known {\it a priori}, detections were defined as a $>3\sigma$ peak at the known location.  The flux uncertainty was measured as the rms in the cleaned portion of the image without known sources.
The peak flux was measured as the highest flux within the detection, while the integrated flux was measured using the routine {\it imfit} and fitting a two-dimensional Gaussian.

\section{Results}

\subsection{Continuum emission}

Table \ref{tab:obs} lists all sources observed at ALMA with the detected Band 6 (1.3 mm; 230 GHz) peak flux and observed rms or 3$\sigma$ limit. For all detected sources, the integrated flux, the beam size and orientation, and the center of the emission are listed.  An integrated flux is not listed if the source was reported as unresolved by {\it imfit}.
The results from {\it imfit} are also used for the reported positions of the detections and to derive the measured component separations given in Table \ref{tab:BinarySep}.  These observations resulted in 21 new detections: 6 single stars, 4 primary stars, and 11 companion stars.

\begin{longrotatetable}
\begin{deluxetable}{llllllllllll}
\tabletypesize{\scriptsize}
\tablewidth{0pt}
\tablecaption{ALMA Observation Results \label{tab:obs}
}
\tablehead{
\colhead{2MASS designation}    & \colhead{Source name} & \colhead{1.3mm Peak} & \colhead{1.3 mm Int.} & \colhead{Beam} &
\colhead{PA} & \colhead{RA} & \colhead{$\sigma_{\rm RA}$} &
\colhead{Dec.} & \colhead{$\sigma_{\rm Dec}$} & \colhead{Deconvolved} & \colhead{Deconvolved}
\\
&  &  \colhead{Flux (mJy)} &  \colhead{Flux (mJy)}   & \colhead{(arcsec)}
& \colhead{(deg)} & \colhead{J2000} & \colhead{(arcsec)}
& \colhead{J2000} & \colhead{(arcsec)} &\colhead{maj. axis}  & \colhead{min. axis}  \\
& & & & & & & & & & \colhead{(mas)} & \colhead{(mas)}
}
\startdata
J04144928+2812305 & FO Tau A &   3.07$\pm$  0.12 &  3.00$\pm$ 0.30 & 0.21x0.14 & 18.9 & 04:14:49.297 & 0.005 & 28:12:30.122 & 0.006 & 161$\pm$ 36 & 126$\pm$ 37 \\ 
J04144928+2812305 & FO Tau B &   2.94$\pm$  0.12 &  3.00$\pm$ 0.30 & 0.21x0.14 & 18.9 & 04:14:49.288 & 0.006 & 28:12:30.086 & 0.007 & ... & ... \\ 
J04183158+2816585 & CZ Tau A & $<$  0.36 & ... & ...& ...& ...& ...& ...& ...  \\ 
J04183158+2816585 & CZ Tau B &   0.60$\pm$  0.12 &  0.62$\pm$ 0.12 & 0.22x0.14 & 19.3 & 04:18:31.621 & 0.018 & 28:16:58.173 & 0.016 & ... & ... \\ 
J04214323+1934133 & IRAS 04187+1927 &   4.24$\pm$  0.10 &  3.72$\pm$ 0.17 & 0.73x0.42 & 51.0 & 04:21:43.243 & 0.040 & 19:34:13.116 & 0.040 & ... & ... \\ 
J04220217+2657304 & FS Tau A &   1.90$\pm$  0.14 &  2.27$\pm$ 0.14 & 0.21x0.14 & 21.3 & 04:22:02.194 & 0.006 & 26:57:30.368 & 0.005 & ... & ... \\ 
J04220217+2657304 & FS Tau B & $<$  0.41 & ... & ...& ...& ...& ...& ...& ...  \\ 
J04263055+2443558 & ... & $<$  0.30 & ... & ...& ...& ...& ...& ...& ...  \\ 
J04295950+2433078 & ... &   3.10$\pm$  0.09 &  2.87$\pm$ 0.15 & 0.87x0.43 & 48.8 & 04:29:59.513 & 0.040 & 24:33:07.285 & 0.040 & 758$\pm$ 20 & 422$\pm$ 8 \\ 
J04300399+1813493 & UX Tau A &  11.50$\pm$  0.50 &  79.00$\pm$ 2.00 & 0.18x0.16 & -174.5 & ... & ... & ... & ... & 470$\pm$ 25 & 320$\pm$ 25 \\ 
J04300399+1813493 & UX Tau Ba & $<$  0.37 & ... & ...& ...& ...& ...& ...& ...  \\ 
J04300399+1813493 & UX Tau Bb & $<$  0.37 & ... & ...& ...& ...& ...& ...& ...  \\ 
J04300399+1813493 & UX Tau C & $<$  0.37 & ... & ...& ...& ...& ...& ...& ...  \\ 
J04302961+2426450 & FX Tau A &   5.60$\pm$  0.12 &  7.84$\pm$ 0.33 & 0.20x0.14 & 24.7 & 04:30:29.659 & 0.003 & 24:26:44.740 & 0.003 & ... & ... \\ 
J04302961+2426450 & FX Tau B & $<$  0.37 & ... & ...& ...& ...& ...& ...& ...  \\ 
J04305137+2442222 & ZZ Tau AB &   0.59$\pm$  0.10 &  0.42$\pm$ 0.13 & 0.80x0.42 & 48.6 & 04:30:51.389 & 0.040 & 24:42:21.864 & 0.040 & ... & ... \\ 
J04314007+1813571 & XZ Tau A &   7.30$\pm$  0.18 &  7.37$\pm$ 0.46 & 0.18x0.16 & -175.5 & 04:31:40.097 & 0.003 & 18:13:56.640 & 0.004 & 93$\pm$ 22 & 51$\pm$ 39 \\ 
J04314007+1813571 & XZ Tau B &   8.70$\pm$  0.18 &  8.92$\pm$ 0.52 & 0.18x0.16 & -175.5 & 04:31:40.082 & 0.003 & 18:13:56.805 & 0.004 & 138$\pm$ 17 & 70$\pm$ 23 \\ 
J04315779+1821380 & V710 Tau A &  53.00$\pm$  0.20 &  66.00$\pm$ 0.56 & 0.73x0.41 & 53.5 & 04:31:57.805 & 0.003 & 18:21:37.616 & 0.003 & 373$\pm$ 12 & 489$\pm$ 2 \\ 
J04315779+1821380 & V710 Tau B & $<$  0.61 & ... & ...& ...& ...& ...& ...& ...  \\ 
J04315968+1821305 & LkHa 267 & $<$  0.30 & ... & ...& ...& ...& ...& ...& ...  \\ 
J04321606+1812464 & BHS98 MHO 5 & $<$  0.31 & ... & ...& ...& ...& ...& ...& ...  \\ 
J04322415+2251083 & ... & $<$  0.30 & ... & ...& ...& ...& ...& ...& ...  \\ 
J04323028+1731303 & GG Tau Aa &   8.70$\pm$  0.74 &  7.05$\pm$ 1.60 & 0.18x0.16 & -173.8 & 04:32:30.364 & 0.009 & 17:31:40.175 & 0.008 & ... & ... \\ 
J04323028+1731303 & GG Tau Ab & $<$  2.22 & ... & ...& ...& ...& ...& ...& ...  \\ 
J04323028+1731303 & GG Tau Ba & $<$  2.40 & ... & ...& ...& ...& ...& ...& ...  \\ 
J04323028+1731303 & GG Tau Bb & $<$  2.40 & ... & ...& ...& ...& ...& ...& ...  \\ 
J04330622+2409339 & GH Tau  A &   3.60$\pm$  0.11 &  3.91$\pm$ 0.20 & 0.20x0.14 & 24.9 & 04:33:06.218 & 0.003 & 24:09:33.640 & 0.003 & ... & ... \\ 
J04330622+2409339 & GH Tau B &   2.60$\pm$  0.11 &  2.89$\pm$ 0.20 & 0.20x0.14 & 24.9 & 04:33:06.239 & 0.004 & 24:09:33.576 & 0.003 & ... & ... \\ 
J04330664+2409549 & V807 Tau A &   8.10$\pm$  0.11 &  8.94$\pm$ 0.26 & 0.20x0.14 & 24.4 & 04:33:06.646 & 0.003 & 24:09:54.737 & 0.003 & ... & ... \\ 
J04330664+2409549 & V807 Tau Bab & $<$  0.33 & ... & ...& ...& ...& ...& ...& ...  \\ 
J04330945+2246487 & ... & $<$  0.28 & ... & ...& ...& ...& ...& ...& ...  \\ 
J04333678+2609492 & IS Tau A &   1.50$\pm$  0.12 &  1.15$\pm$ 0.12 & 0.20x0.14 & 24.4 & 04:33:36.804 & 0.007 & 26:09:48.777 & 0.008 & ... & ... \\ 
J04333678+2609492 & IS Tau B &   1.20$\pm$  0.12 &  1.05$\pm$ 0.12 & 0.20x0.14 & 24.4 & 04:33:36.816 & 0.009 & 26:09:48.663 & 0.011 & ... & ... \\ 
J04333935+1751523 & HN Tau A &   7.10$\pm$  0.10 &  15.70$\pm$ 1.90 & 0.72x0.41 & 53.6 & 04:33:39.376 & 0.033 & 17:51:51.974 & 0.042 & 1390$\pm$ 120 & 350$\pm$ 140 \\ 
J04333935+1751523 & HN Tau B &   0.57$\pm$  0.10 &  0.54$\pm$ 0.10 & 0.72x0.41 & 53.6 & 04:33:39.231 & 1.048 & 17:51:49.716 & 1.382 & ... & ... \\ 
J04355684+2254360 & Haro 6-28 A &   4.90$\pm$  0.10 &  5.14$\pm$ 0.20 & 0.19x0.16 & -178.4 & 04:35:56.865 & 0.003 & 22:54:35.805 & 0.003 & 91$\pm$ 14 & 50$\pm$ 42 \\ 
J04355684+2254360 & Haro 6-28 B &   1.05$\pm$  0.10 &  0.78$\pm$ 0.15 & 0.19x0.16 & -178.4 & 04:35:56.822 & 0.006 & 22:54:35.539 & 0.009 & ... & ... \\ 
J04361030+2159364 & ... & $<$  0.31 & ... & ...& ...& ...& ...& ...& ...  \\ 
J04362151+2351165 & ... &   0.15$\pm$  0.10 &  1.45$\pm$ 0.19 & 0.79x0.42 & 48.9 & 04:36:21.508 & 0.040 & 23:51:16.300 & 0.040 & ... & ... \\ 
J04391741+2247533 & VY Tau A &   1.28$\pm$  0.10 &  1.95$\pm$ 0.27 & 0.19x0.16 & 3.4 & 04:39:17.429 & 0.008 & 22:47:53.044 & 0.010 & 177$\pm$ 45 & 96$\pm$ 70 \\ 
J04391741+2247533 & VY Tau B & $<$  0.31 & ... & ...& ...& ...& ...& ...& ...  \\ 
J04392090+2545021 & GN Tau A &   0.62$\pm$  0.11 &  0.47$\pm$ 0.11 & 0.20x0.15 & -179.0 & 04:39:20.912 & 0.013 & 25:45:01.820 & 0.013 & ... & ... \\ 
J04392090+2545021 & GN Tau B &   0.64$\pm$  0.11 &  0.59$\pm$ 0.11 & 0.20x0.15 & -179.0 & 04:39:20.938 & 0.012 & 25:45:01.525 & 0.010 & ... & ... \\ 
J04404950+2551191 & JH 223 A &   1.10$\pm$  0.12 &  1.68$\pm$ 0.30 & 0.24x0.13 & 34.9 & 04:40:49.516 & 0.011 & 25:51:18.662 & 0.011 & 172$\pm$ 52 & 52$\pm$ 62 \\ 
J04404950+2551191 & JH 223 B &   0.77$\pm$  0.12 &  0.76$\pm$ 0.20 & 0.24x0.13 & 34.9 & 04:40:49.466 & 0.009 & 25:51:20.695 & 0.018 & ... & ... \\ 
J04410826+2556074 & ITG 33A &   1.90$\pm$  0.11 &  4.10$\pm$ 0.38 & 0.24x0.13 & 34.8 & 04:41:08.271 & 0.005 & 25:56:07.033 & 0.010 & 279$\pm$ 36 & 141$\pm$ 32 \\ 
J04411078+2555116 & ITG 34 &   0.84$\pm$  0.11 &  0.70$\pm$ 0.11 & 0.24x0.13 & 34.8 & 04:41:10.794 & 0.010 & 25:55:11.228 & 0.010 & ... & ... \\ 
J04412464+2543530 & ITG 40 &   0.85$\pm$  0.11 &  1.00$\pm$ 0.12 & 0.24x0.13 & 34.8 & 04:41:24.661 & 0.016 & 25:43:52.608 & 0.009 & ... & ... \\ 
J04414489+2301513 & ... & $<$  0.38 & ... & ...& ...& ...& ...& ...& ...  \\ 
J04420777+2523118 & V955 Tau A &   1.80$\pm$  0.11 &  2.16$\pm$ 0.24 & 0.21x0.16 & -174.5 & 04:42:07.787 & 0.005 & 25:23:11.580 & 0.007 & 105$\pm$ 41 & 97$\pm$ 80 \\ 
J04420777+2523118 & V955 Tau B &   0.86$\pm$  0.11 &  0.86$\pm$ 0.21 & 0.21x0.16 & -174.5 & 04:42:07.770 & 0.009 & 25:23:11.201 & 0.014 & 105$\pm$ 71 & 25$\pm$ 84 \\ 
J04423769+2515374 & DP Tau A &   2.10$\pm$  0.11 &  2.10$\pm$ 0.35 & 0.21x0.17 & -40.0 & 04:42:37.696 & 0.012 & 25:15:36.924 & 0.009 & 246$\pm$ 40 & 102$\pm$ 66 \\ 
J04423769+2515374 & DP Tau B &   1.50$\pm$  0.11 &  1.50$\pm$ 0.20 & 0.21x0.17 & -40.0 & 04:42:37.693 & 0.041 & 25:15:37.124 & 0.029 & ... & ... \\ 
J04432023+2940060 & ... & $<$  0.38 & ... & ...& ...& ...& ...& ...& ...  \\ 
J04465897+1702381 & Haro 6-37 A &   0.80$\pm$  0.33 &  2.10$\pm$ 0.90 & 0.18x0.16 & 9.8 & 04:46:58.975 & 0.022 & 17:02:37.631 & 0.020 & 143$\pm$ 64 & 86$\pm$ 95 \\ 
J04465897+1702381 & Haro 6-37 B & $<$  0.97 & ... & ...& ...& ...& ...& ...& ...  \\ 
J04465897+1702381 & Haro 6-37 C &  13.40$\pm$  0.33 &  38.50$\pm$ 1.90 & 0.18x0.16 & 9.8 & 04:46:59.090 & 0.005 & 17:02:39.713 & 0.005 & 357$\pm$ 16 & 289$\pm$ 14 \\ 
J04554535+3019389 & ... & $<$  0.39 & ... & ...& ...& ...& ...& ...& ...  \\ 
J04554801+3028050 & ... & $<$  0.35 & ... & ...& ...& ...& ...& ...& ...  \\ 
J04554969+3019400 & ... & $<$  0.35 & ... & ...& ...& ...& ...& ...& ...  \\ 
J04560118+3026348 & XEST 26-071 & $<$  0.35 & ... & ...& ...& ...& ...& ...& ...  \\ 
J05052286+2531312 & CIDA 9 A &   7.20$\pm$  0.14 &  33.80$\pm$ 0.80 & 0.25x0.16 & 44.0 & ... & ... & ... & ... & 380$\pm$ 25 & 300$\pm$ 25 \\ 
J05052286+2531312 & CIDA 9 B & $<$  0.41 & ... & ...& ...& ...& ...& ...& ...  \\ 
J05062332+2432199 & CIDA 11 & $<$  0.36 & ... & ...& ...& ...& ...& ...& ...  \\

\enddata
\tablecomments{The deconvolved sizes given for UX Tau A and CIDA 9 A correspond to the major and minor axis of the peak flux, which is a ring in both cases.}
\end{deluxetable}
\end{longrotatetable}

\begin{figure}
    \centering
    \includegraphics[width=4.1in]{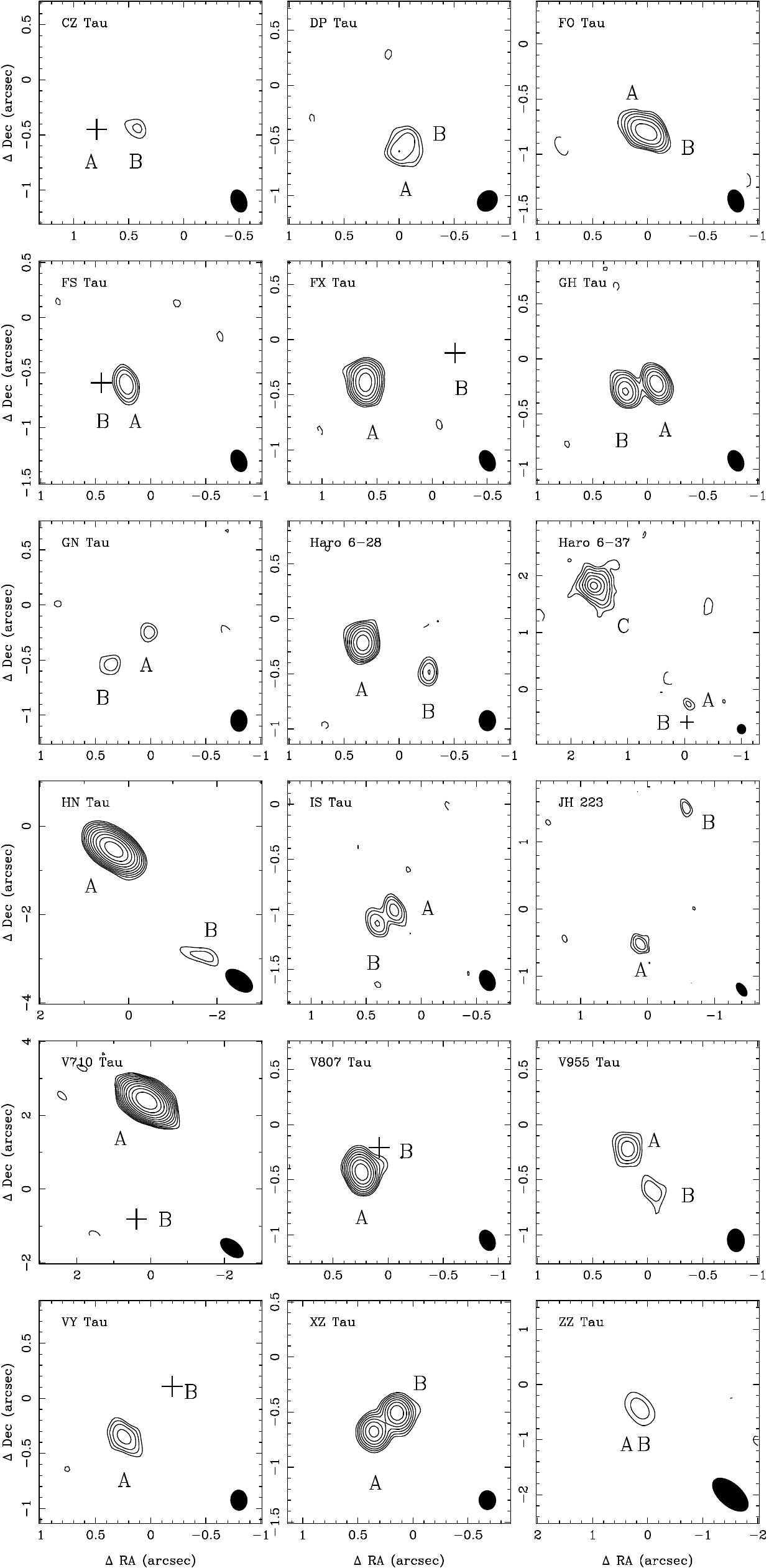}
   \caption{230 GHz continuum images for all multiple systems in which the individual components are unresolved or Gaussian. Note that the spatial scale varies depending on the beam size and component separation. Contours start at 3$\sigma$, as reported in Table \ref{tab:obs}, and increase by 50\% in each step. }
    \label{fig:multis}
\end{figure}

\begin{figure}
    \centering
    \includegraphics[width=4.5in]{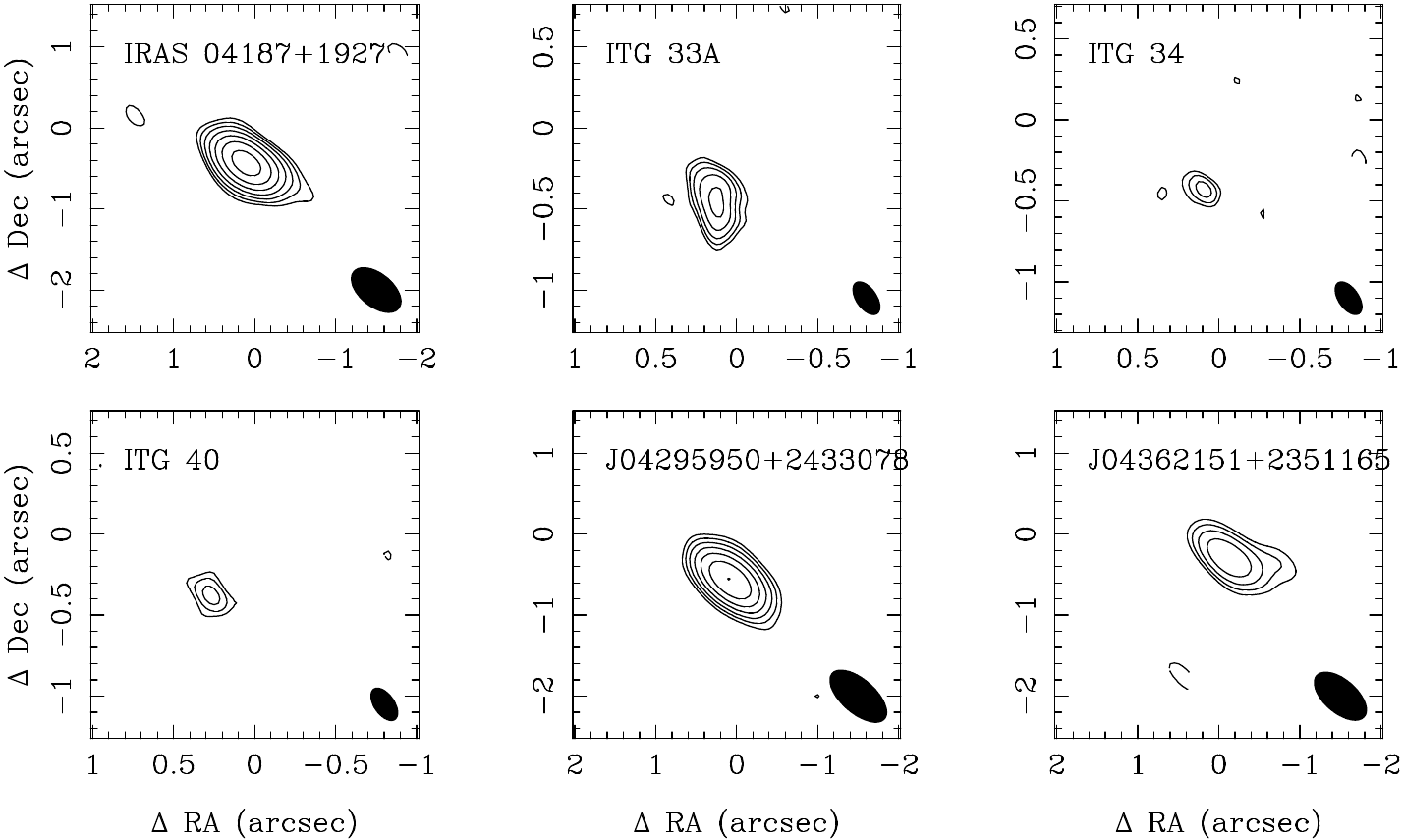}
    \caption{230 GHz continuum images for detected single stars.  Contours start at 3$\sigma$, as reported in Table \ref{tab:obs},  and increase by 50\% in each step.}
    \label{fig:singles}
\end{figure}

Figure \ref{fig:multis} plots the continuum emission for the multiple systems where the emission is unresolved or well fit by a Gaussian.
In every multiple system observed, we detected continuum emission from at least one disk.  The positions of the components are marked in the case of non-detections.  In some cases, older catalog coordinates were used to set the ALMA pointing, resulting in the offsets seen in Figure \ref{fig:multis}.  We
have confirmed that our detected positions correspond to the expected stellar positions, and the differences due to orbital motion are discussed below.
Figure \ref{fig:singles} plots the continuum emission from all detected single stars.

\begin{figure}
    \centering
    \includegraphics[width=4in]{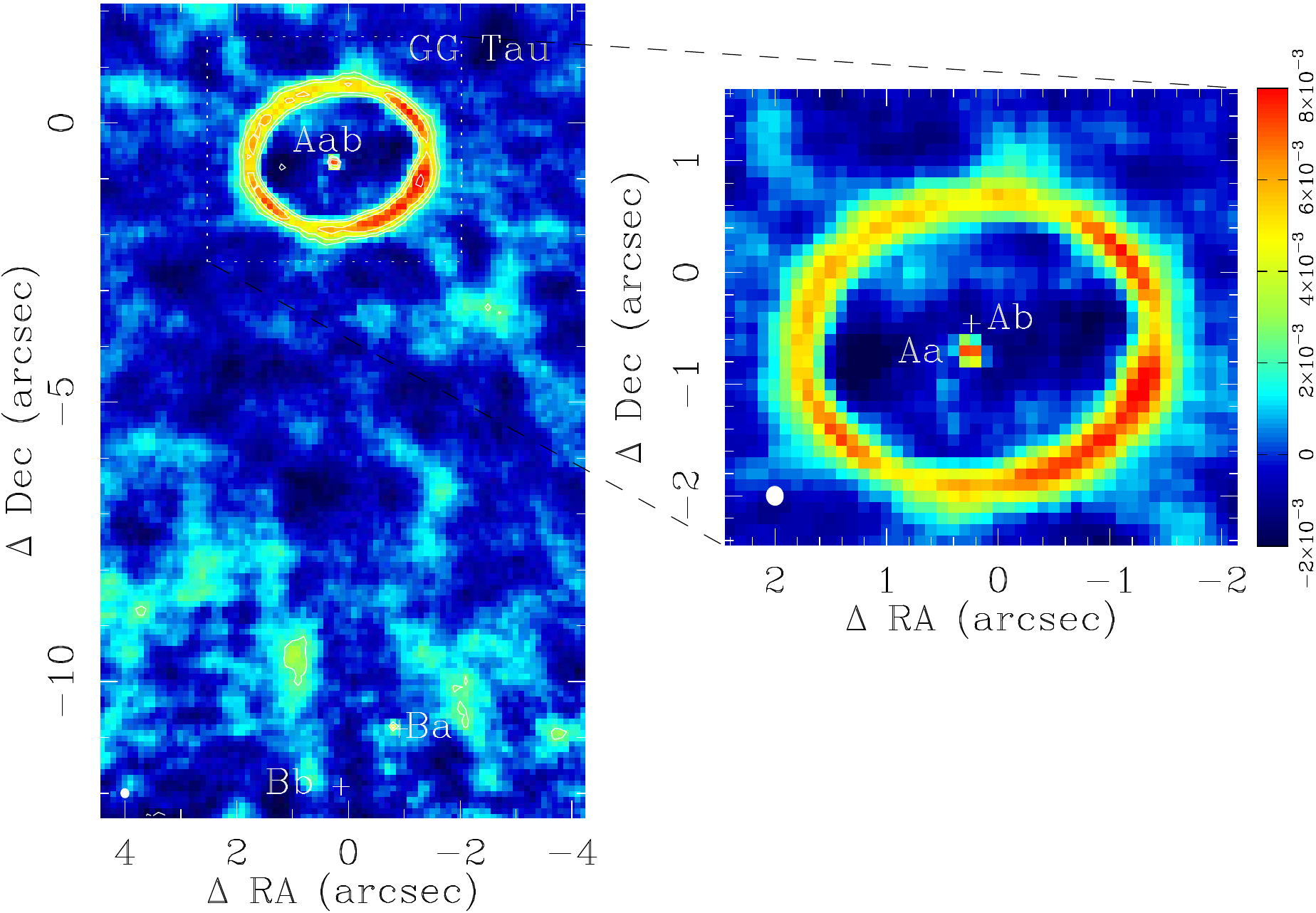}
    \caption{230 GHz continuum image for GG Tau.  The units for the color scale are Jy$\times$beam$^{-1}$. Contours start at 3$\sigma$ and increase by 50\% in each step.}
    \label{fig:ggtau}
\end{figure}

\begin{figure}
    \centering
    \includegraphics[width=3in]{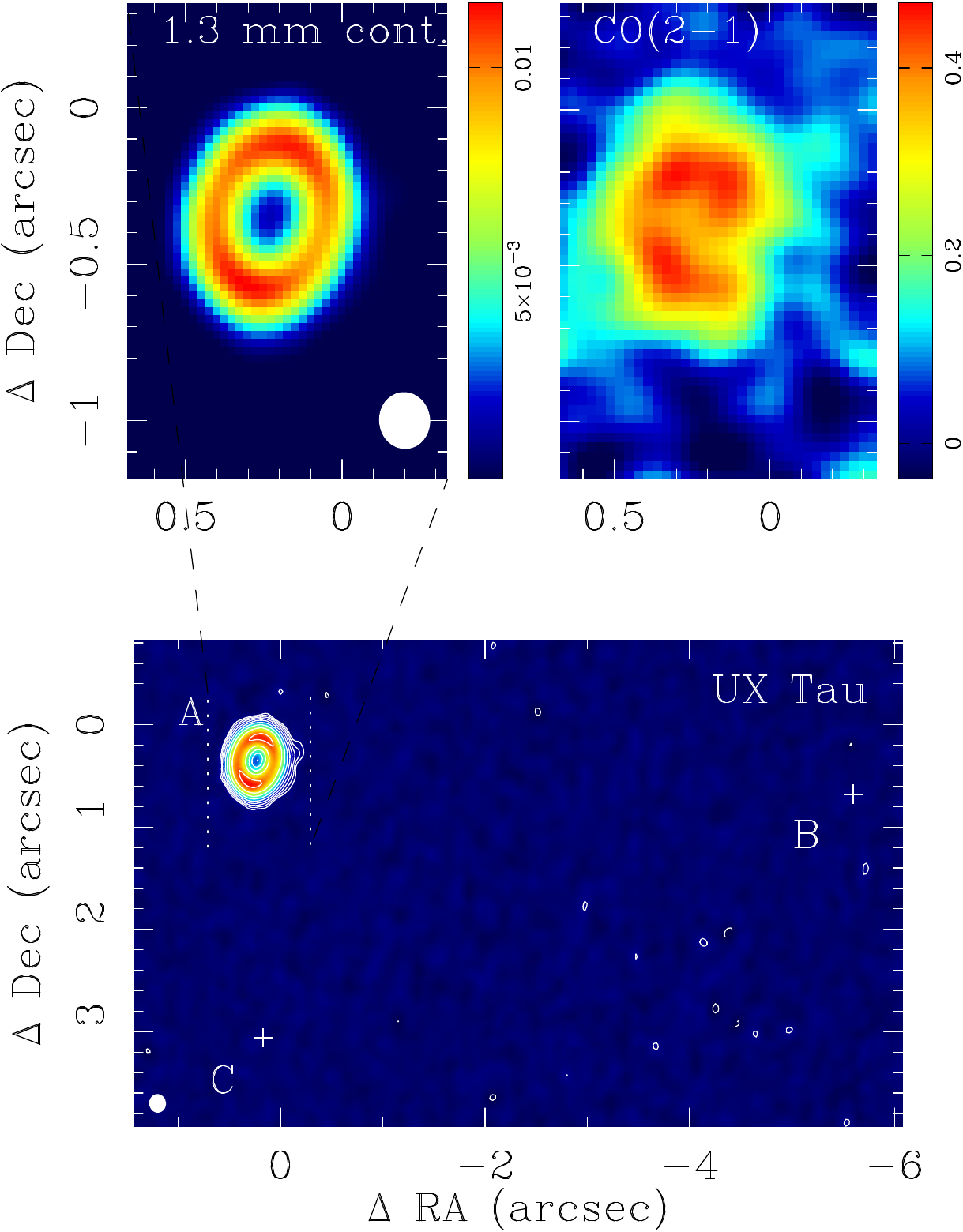}
    \caption{Bottom and upper left panels: 230 GHz continuum image for UX Tau. The units for the color scale are Jy$\times$beam$^{-1}$. Contours start at 3$\sigma$ and increase by 50\% in each step.  Upper right panel:  The CO (2-1) map integrated over all velocities (moment 0) for the UX Tau inset panel.  The units for the color scale are Jy$\times$beam$^{-1}\times$km$\times$sec$^{-1}$.}
    \label{fig:uxtau}
\end{figure}

\begin{figure}
    \centering
    \includegraphics[width=3in]{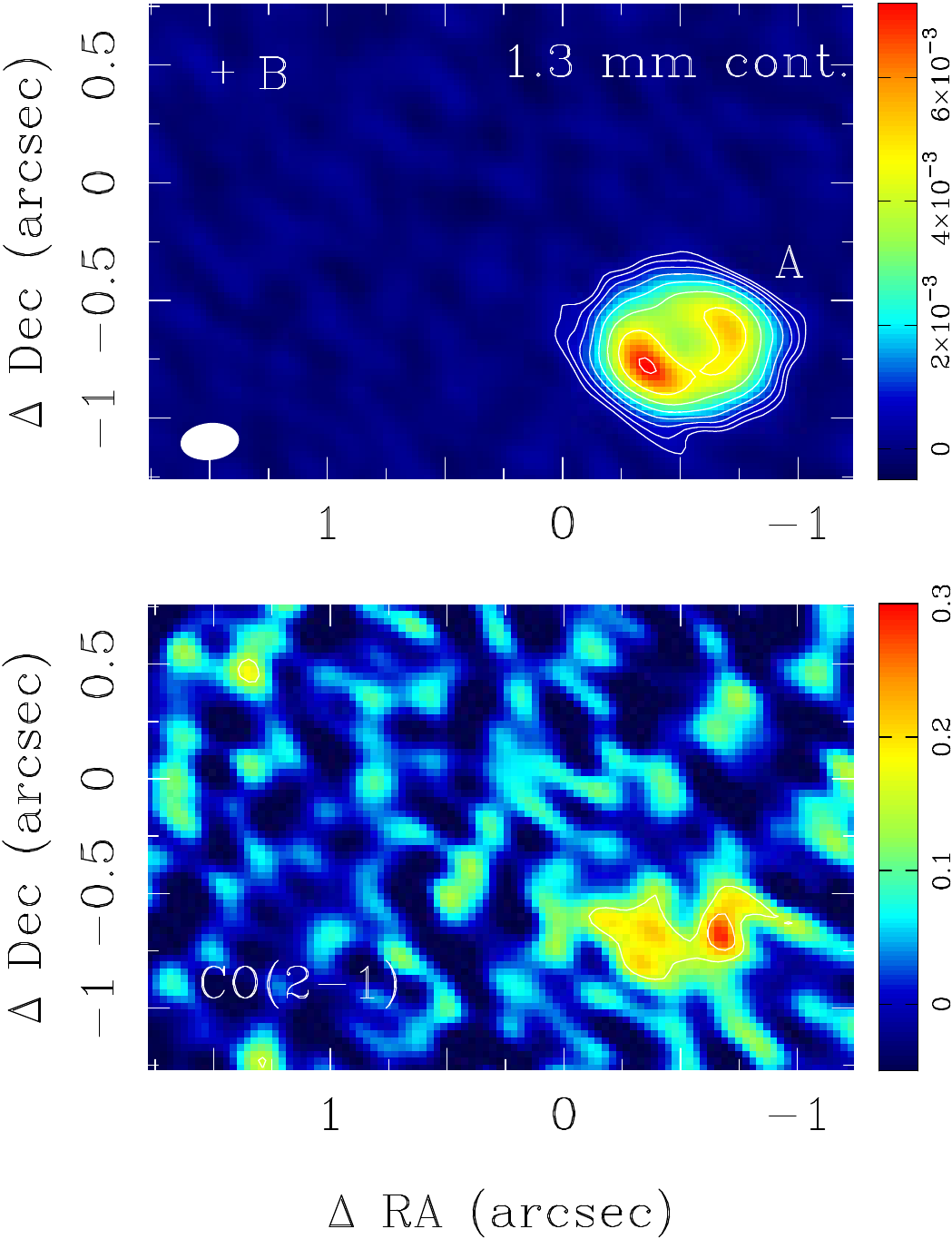}
    \caption{Top: 230 GHz continuum image for CIDA 9. The units for the color scale are Jy$\times$beam$^{-1}$. Contours start at 3$\sigma$ and increase by 50\% in each step. Bottom: The CO (2-1) integrated velocity (moment 0) map for CIDA 9.  The units for the color scale are Jy$\times$beam$^{-1}\times$km$\times$sec$^{-1}$.}
    \label{fig:cida9}
\end{figure}

For three of the sources, our high angular resolution observations trace structure in the disk.  Figure \ref{fig:ggtau} shows the well-studied circumbinary ring of GG Tau.  Note that we can localize the central emission to arise from a circumstellar disk around GG Tau Aa.  
UX Tau A was identified as a pre-transition disk by \citet{esp07} using {\it Spitzer} data, and our ALMA data show a strong clearing in the inner disk
(Figure \ref{fig:uxtau}).  From the mid-infrared spectra, CIDA 9 was not identified as a transition disk by \citet{fur11}, but the ALMA data show a ring-like structure with an emission deficit in the center (Figure \ref{fig:cida9}).   

For 11 of the binary or multiple star systems, we detected emission from two components in the system and thus are able to measure projected separations and position angles, shown in Table \ref{tab:BinarySep}.  Although our positions measure the centroid of the millimeter-wavelength emission rather than the stellar position, comparison of our measured positions with those from Gaia DR2 \citep{gaia18} typically agree to within a few tens of milliarcseconds, comparable to the ALMA astrometric uncertainty for our data, indicating that our measured positions trace the stellar positions very well and that the millimeter emission is not substantially asymmetric or offset from the stars at our resolution.  As such, we can compare our measured separations and position angles with those in the literature from earlier epochs.  The four widest sources, HN Tau AB, JH 223 AB, Haro 6-28 AB, and Haro 6-37 AC, all have position angles and projected separations that agree with previous observations (Table \ref{tab:BinarySep}).  In contrast, all of the sources with projected separations of 0\farcs5 or less ($\sim 70$ AU at an assumed distance of 140 pc) show detectable orbital motion.  In some cases the motion is substantial (e.g.\ a 60\arcdeg\ change in position angle for FO Tau), but in all of these cases it is within the amount of motion expected from simple assumptions about the orbits (e.g.\ modest eccentricities and that the semimajor axis is similar to the current projected separation).  Many of these systems had tentative initial detections of orbital motion in \citet{woi01} and our results show continued motion consistent with their results.  Perhaps the most surprising system is Haro 6-28 AB, which with a projected separation of 0\farcs65 might have been expected to show orbital motion, but which has a separation and position angle consistent with that measured by \citet{whi01} in 1997, suggesting that it may be near periastron in an eccentric orbit and/or that the true separation may be substantially larger than the projected separation.   We also note that given the few tens of mas yr$^{-1}$ proper motions of typical sources in Taurus, comparison with previous measurements shows that all of the sources in Table \ref{tab:BinarySep} are common-proper-motion systems. 

\begin{deluxetable}{rlllll}
\tablewidth{0pt}
\tablecaption{Measured component separations \label{tab:BinarySep}
}
\tablehead{
\colhead{Name}     & \colhead{ALMA separation} & \colhead{ALMA PA} & \colhead{Lit. separation} & \colhead{Lit. PA} & \colhead{Lit. reference} \\
& \colhead{(arcsec)}   & \colhead{(deg)} & \colhead{(arcsec)}  & \colhead{(deg)} \\
}
\startdata
FO Tau AB &  0.124$\pm$0.012 &   253.2$\pm$3.7  & 0.150$\pm$0.007 &   193.7$\pm$1.0  & White and Ghez (2001)  \\ 
XZ Tau AB &  0.271$\pm$0.006 &   307.4$\pm$0.9  & 0.300$\pm$0.006 &   324.5$\pm$1.0  & White and Ghez (2001)  \\ 
GH Tau  AB &  0.287$\pm$0.007 &   103.0$\pm$1.0  & 0.305$\pm$0.006 &   114.8$\pm$1.1  & White and Ghez (2001)  \\ 
IS Tau AB &  0.195$\pm$0.017 &   125.9$\pm$3.5  & 0.222$\pm$0.004 &    95.4$\pm$1.4  & White and Ghez (2001)  \\ 
HN Tau AB &  3.061$\pm$1.735 &   222.5$\pm$23.2 & 3.142$\pm$0.001 &   219.7$\pm$0.5  & Correia et al (2006)  \\ 
Haro 6-28 AB &  0.651$\pm$0.012 &   245.8$\pm$0.6  & 0.647$\pm$0.012 &   245.2$\pm$1.0  & White and Ghez (2001)  \\ 
GN Tau AB &  0.450$\pm$0.024 &   130.9$\pm$2.1  & 0.335$\pm$0.006 &   124.1$\pm$1.0  & White and Ghez (2001)  \\ 
JH 223 AB &  2.145$\pm$0.026 &   341.4$\pm$0.6  & 2.060$\pm$0.100 &   342.3$\pm$2.8  & Kraus and Hillenbrand (2007)  \\ 
V955 Tau AB &  0.447$\pm$0.019 &   212.1$\pm$1.8  & 0.323$\pm$0.008 &   204.0$\pm$1.2  & White and Ghez (2001)  \\ 
DP Tau AB &  0.207$\pm$0.052 &   345.7$\pm$8.6  & 0.107$\pm$0.001 &   293.3$\pm$0.3  & Kraus et al (2011)  \\ 
Haro 6-37 AC &  2.660$\pm$0.031 &    38.5$\pm$0.5  & 2.650$\pm$0.080 &    38.9$\pm$0.2  & Schaefer et al (2014)  \\ 

\enddata
\end{deluxetable}

To determine disk masses, we need to know the distance to each of our sources.  Gaia DR2 \citep{gaia18} contains parallax measurements for most of our sources, but since many of our sources are binaries, we need to be careful that the astrometric solution is not affected by orbital acceleration in the system.  To assess the quality of each astrometric solution, we followed the procedure recommended in Gaia Technical Note GAIA-C3-TN-LU-LL-124-01 \citep{Lindegren2018}.  For each star with a measured Gaia parallax, we calculated the renormalized unit weight error (RUWE) of the astrometric solution, retaining only those with RUWE $\le 1.6$.  For sources with good astrometric solutions, we took distances and uncertainties from \citet{BailerJones2018}. In all cases, we used the distance of the brightest Gaia source within 8\arcsec\ of the 2MASS position; in particular, this means that we used the same distance for all components in a given binary or multiple system.  The lone exception is the pair GI and GK Tau, which are separated by 13\farcs2.   For sources without reliable Gaia DR2 distances, we used the weighted mean of the Gaia distances for all other sources in our sample with 30\arcmin.  A small number of sources had no sources in our sample within 30\arcmin\ that had reliable distances; in those cases, we adopted the median distance from our sample of 138.8$\pm18.8$ pc. 

\begin{figure}
    \centering
    \includegraphics[width=5.5in]{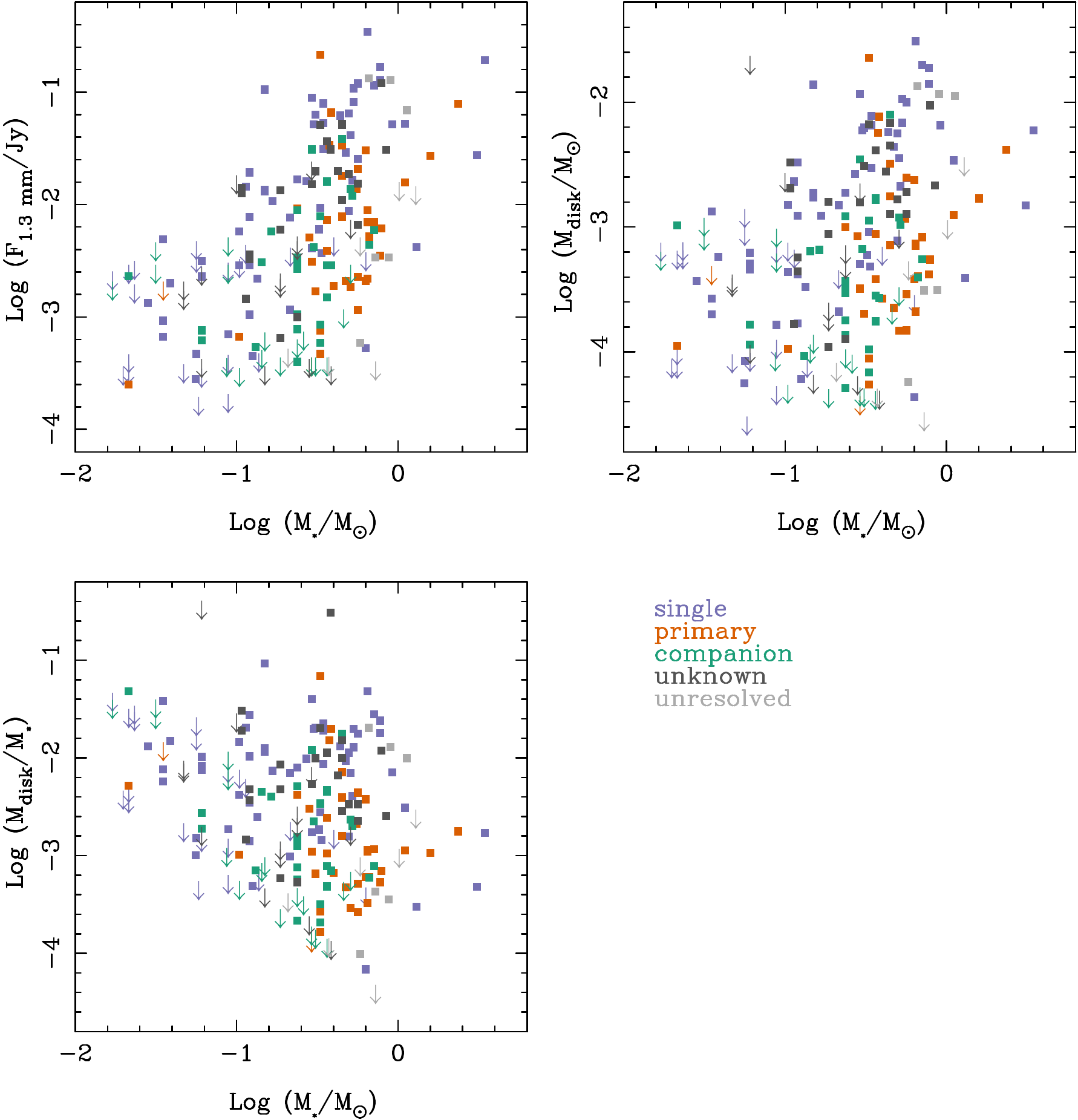}
    \caption{Top left: The 1.3 mm flux plotted against the stellar mass for all Class II stars listed in Table \ref{tab:sample}.  Top right: The calculated disk mass plotted against the stellar mass for the same stars. Bottom left: The ratio of the disk to stellar mass plotted against the stellar mass for the same stars.}
    \label{fig:flux}
\end{figure}

Assuming the dust is optically thin, the conversion from flux ($F_{\nu}$) to
disk mass ($M_d$) is
\begin{equation}
M_d = \frac{F_{\nu}d^2}{\kappa_{\nu}X_gB_{\nu}(T_d)}.
\end{equation}
For comparison to the Taurus sample results of \citet{and13}, we use the same constants of
dust-to-gas ratio $X_g$ = 0.01 and dust opacity $\kappa_{\nu}$ = 2.3 cm$^2$ g$^{-1}$ at 1.3 mm.  For our new ALMA sources, the uncertainty listed in Table \ref{tab:sample} includes both the observed uncertainty
from Table \ref{tab:obs} and a 5\% absolute flux calibration uncertainty (ALMA memo 594).  For the
mean dust temperature $T_d$, we also adopt the \citet{and13} scaling of $T_d = 25 (L_{\ast}/L_{\odot})^{-1/4}$~K.  Our derivation of the stellar luminosity is described in \S \ref{sec:stellar} and the calculated dust temperatures range from 7 to 67 K, but the dust temperature for 95\% of the single and binary star sample defined in \S \ref{sec:sample} ranges from 10 to 30 K.  The derived disk mass or limit is given in Table \ref{tab:sample}(at end of paper)  and shown for all stars in Figure \ref{fig:flux}.

For most of our ALMA data, the 3$\sigma$ non-detection limit is
$\sim$0.35 mJy which corresponds to a total disk mass limit of $4 \times 10^{-5}$ M$_{\odot}$ = $4 \times 10^{-2}$ M$_{\rm Jup} \approx M_{\rm Uranus}$ for a spectral type of M3.
Even at this sensitivity level, we have detected no new circumbinary disks, suggesting that circumbinary disks are not a common outcome of the binary star formation process or that they dissipate by an age of 1--2 Myr.

\subsection{CO emission}

Because our observing strategy was tailored for a large survey and focused on detecting continuum emission, our exposure times are not sufficient to detect CO(2--1) emission from the faint disks that form the majority of our new detections.  Of 26 systems where we detect continuum emission (\S \ref{tab:obs}), we detect CO from only 16 systems, and then usually from only one component in the system, so we are unable to do any meaningful statistical analysis on the CO emission.  Nonetheless, since a fraction of our systems do show some evidence of CO emission, in Appendix A we provide some brief notes about the detected sources, and we provide peak fluxes and RMS values from the integrated CO emission in Table \ref{tab:CO}. 

\subsection{Stellar properties}
\label{sec:stellar}

In order to compare the results from the ALMA survey with the population of Class II stars in Taurus, we need stellar properties for all 211 of the Class II stars.
For consistency, we followed the same procedure for all stars, even those not observed by ALMA.
First, we compiled spectral types from the literature, trying wherever possible to go to the original references (listed in Table \ref{tab:sample}) rather than adopting spectral types referenced in other compilations.  If the only available reference for a star's spectral type gave a range of types, we adopted the middle of that range, e.g. for a spectral type given as M3--M5 we adopted M4.  Using these spectral types, we followed the approach of \citet{wd18} to determine stellar masses and luminosities.  More specifically, we converted spectral types to effective temperatures using the scale from \citet{HH2014}. Two of our sources have spectral type M9.25, while the \citet{HH2014} effective temperature scale stops at M9.  For these sources, we adopted a simple linear extrapolation of the M7--M9 temperatures, resulting in an adopted effective temperature of 2545 K for these two sources, 25 K cooler than M9.  For all sources, we adopted an assumed spectral type uncertainty of $\pm1$ spectral subclass, and translated that into an uncertainty on effective temperature.  We then adopted an assumed age of 1 Myr for Taurus, consistent with the finding of \citet{Kraus2009b} that 1--2 Myr is a good fit to the cluster sequence.  We used that age and the effective temperature to determine a stellar mass and luminosity for each star from the models of \citet{Baraffe2015} for $T_{\rm eff} \le 4210$ K and the MIST models from \citet{Choi2016} for $T_{\rm eff} > 4210$ K\null.  In all cases, we used cubic spline interpolation between tabulated values, and propagated the spectral type uncertainty into the uncertainties we quote for $M_{\star}$ and $L_{\star}$ (Table \ref{tab:sample}).

\section{Analysis}

\subsection{Defining a sample for statistical analysis}
\label{sec:sample}

As noted above, we included only systems categorized as Class II in assembling the target list for our survey.  For binary and multiple systems, the system is included if only one classification is available, as is the case for most close binaries, but we rejected systems where one component has been clearly classified as Class I or Class III.  The only multiple systems eliminated by these criteria are T Tau (triple), V773 Tau (triple), UX Tau (quadruple), and 2MASS J04554757+3028077/2MASS J04554801+3028050, a 6\arcsec\ binary where the primary is a Class III star and both components have no millimeter detection.
UX Tau was excluded after the observations were taken, so its data are presented in Tables \ref{tab:obs} and \ref{tab:sample}. 
To rigorously define a sample for statistical analysis, we
consider stellar mass and multiplicity status.  The detailed criteria and cutoffs for these categories are described in the sub-sections below, and Table \ref{tab:counts} gives
the counts and detection fractions.  To standardize our comparisons and analysis, we have assembled
fluxes, derived stellar masses and luminosities (\S \ref{sec:stellar}), and calculated disk masses for all Class II objects in Taurus, including those that are not included in our statistical sample (Table \ref{tab:sample}).  The flux values come primarily from this work and from the compilation of \citet{and13}, supplemented by more recent ALMA data or limits where available \citep[Paper 1]{har15,wd18}.

\subsection{Stellar Mass}

Although our input target list was complete in Class II systems to a spectral type of M8, the fact that one of our requested observing blocks was not observed resulted in a substantial number of lower mass single stars still having 1.3 mm flux upper limits of 1 to $\sim$20 mJy, significantly worse than the ALMA detections and limits.
To avoid biased conclusions from these upper limits, we have defined a stellar mass cutoff above which at least 75\% of the stars have been detected.  For both single and binary stars, this stellar mass cutoff is 0.06 M$_{\odot}$ or a spectral type of M6.  In practice, the only binary systems that this eliminates from the statistical sample are FU Tau, an isolated pair of brown dwarfs on the edge of the Taurus cloud \citep{luh10} and DH Tau, a system where the companion may be of planetary mass \citep{Itoh2005, Luhman2006c, Bonnefoy2014}.  We did not observe FU Tau with ALMA, and the millimeter upper limits from previous observations of this system are not particularly sensitive,  
especially given the very low mass of the pair.   However, all new observations, regardless of stellar mass, are listed in Table \ref{tab:obs} and Table \ref{tab:sample}.

\subsubsection{Single star sample}
In order to isolate the effects of multiplicity on disk properties as cleanly as possible, it is essential to define our single star comparison sample carefully.  Past practice (including in some of our own work) has often been to assume that stars not previously shown to be binary could be treated as single.  However, high resolution observations of T Tauri stars have continued to uncover previously unknown companions, particularly at smaller separations and/or lower masses than previously detected.  

Thus, we have taken extra care in this work to define our comparison sample of single stars.  For a given star to be included in the single star sample, we require that there are published observations that were sensitive to stellar companions with separations as close as 20 AU (roughly 0\farcs14 at the distance of Taurus). We drew limits on companions primarily from \cite{kra11, kra12}, who give limits on companions out to 30\arcsec, with a handful of systems from other papers.  While it is possible that there remain a small number of undetected companions in our single star sample, we are confident that it is largely free of stellar mass companions, especially those expected to have a large effect on disks.  

In addition to the high-resolution imaging observations, we require that there is a published millimeter flux or upper limit, which was available for all sources.  After applying the stellar mass limits discussed above, our final sample of single stars consists of 65 stars, designated with a ``0'' in the ``Role'' column and a ``Y'' in the Sample column in Table \ref{tab:sample}.

\subsubsection{Binary and Multiple star sample}
\label{sec:binary}

For our analysis of the effects of multiplicity, we restrict the binary sample to stars that are strictly binary (only two stars in the system) and that could be resolved by the spatial resolution of these observations ($>0\farcs18$), excluding triples and higher-order multiples to avoid the ambiguity of the influence of both close and wide pairs in the same system.  For this binary statistical sample, we require that both stars in the system have been observed with high spatial resolution, applying the same criterion as for the single-star sample above.  The outer cutoff was set at a projected separation of 30\arcsec\ (4200 AU) following \citet{kra09} as this is the separation in Taurus at which the frequency of chance alignments becomes significant.
This results in a sample of 30 binaries.  In Table \ref{tab:sample} the primaries are designated with a ``1'' and the secondaries with a ``2'' in the Role column, and those in the statistical sample with a ``Y'' in the Sample column.

For the binary comparisons, we define the primary star as the star with the highest stellar mass.
To divide the binary sample into wide and close pairs, we used a projected separation value of 1\arcsec\ or $\sim140$ AU.  This is based partly on surveys that show a typical continuum disk size of $\sim$50--100 AU but also results in a reasonable number of systems in each sub-sample for statistical comparison.  
The total number of stars and the number of detections in each of these categories (single, wide binary, close binary) are given
in Table \ref{tab:counts}.  Within this sample, all
detection fractions are above 75\%, even for the closest binaries ($0\farcs11$ to $0\farcs4$ or 15 to 60 AU). 

\begin{deluxetable}{llll}
\tabletypesize{\scriptsize}
\tablewidth{0pt}
\tablecaption{Sample Size and Detection Rates \label{tab:counts}
}
\tablehead{
\colhead{Sample} & \colhead{Total}   & \colhead{mm detections} & \colhead{Detection percentage} 
}
\tablenum{5}
\startdata
All Class II & 211 & 151 & $72\%$ \\
M6 or earlier  & 189 & 143 & $76\%$ \\
Singles & 65 & 52 & $80\%$ \\
Binaries \\
\quad All primaries & 30 & 29 & $97\%$ \\ 
\quad All secondaries & 30 & 24 & $80\%$ \\
\quad Wide primaries & 12 & 12 & $100\%$ \\
\quad Wide secondaries & 12 & 10 & $83\%$ \\
\quad Close primaries & 18 & 17 & $94\%$ \\ 
\quad Close secondaries & 18 & 14 & $78\%$ \\
\enddata
\end{deluxetable}

\subsection{Statistical Calculations}

After applying the definitions of single and binary given above, as well as the stellar mass criteria, we have a sample of 125 stars: 65 single stars, and 60 stars in 30 binary pairs.  The millimeter flux detection fraction for this sample is 80\% or higher for each category of single, primary, or secondary.  This detection fraction is substantially higher than for previous surveys; by comparison, the detection fraction above M6 in \citet{and13} was 58\% and the detection fraction in multiple systems in \citet{har12} was 28--37\%.  This detection efficiency gives us more fidelity in comparing the influence of binarity on the millimeter flux and therefore on the disk mass.  

To quantitatively compare the properties of the samples, we used the R statistical package, in all cases using routines for censored data to account for the upper limits in millimeter flux.  The results from these statistical analyses are given in the following sections.  From the {\it survival\/} package 
we used the log rank test from the {\it survdiff\/} function for two-sample comparisons, and the Kaplan-Meier (KM) maximum likelihood estimator in the {\it survfit\/} function to compare the distribution functions.  We used the {\it cenken\/} function to calculate the Kendall's $\tau$ correlation coefficient between the primary and secondary disk to stellar mass ratios, and to estimate the slope of any correlation.

\subsection{Accounting for stellar mass in comparing the samples}
\label{sec:stellarmass}

\begin{figure}
    \centering
    \includegraphics[width=5.5in]{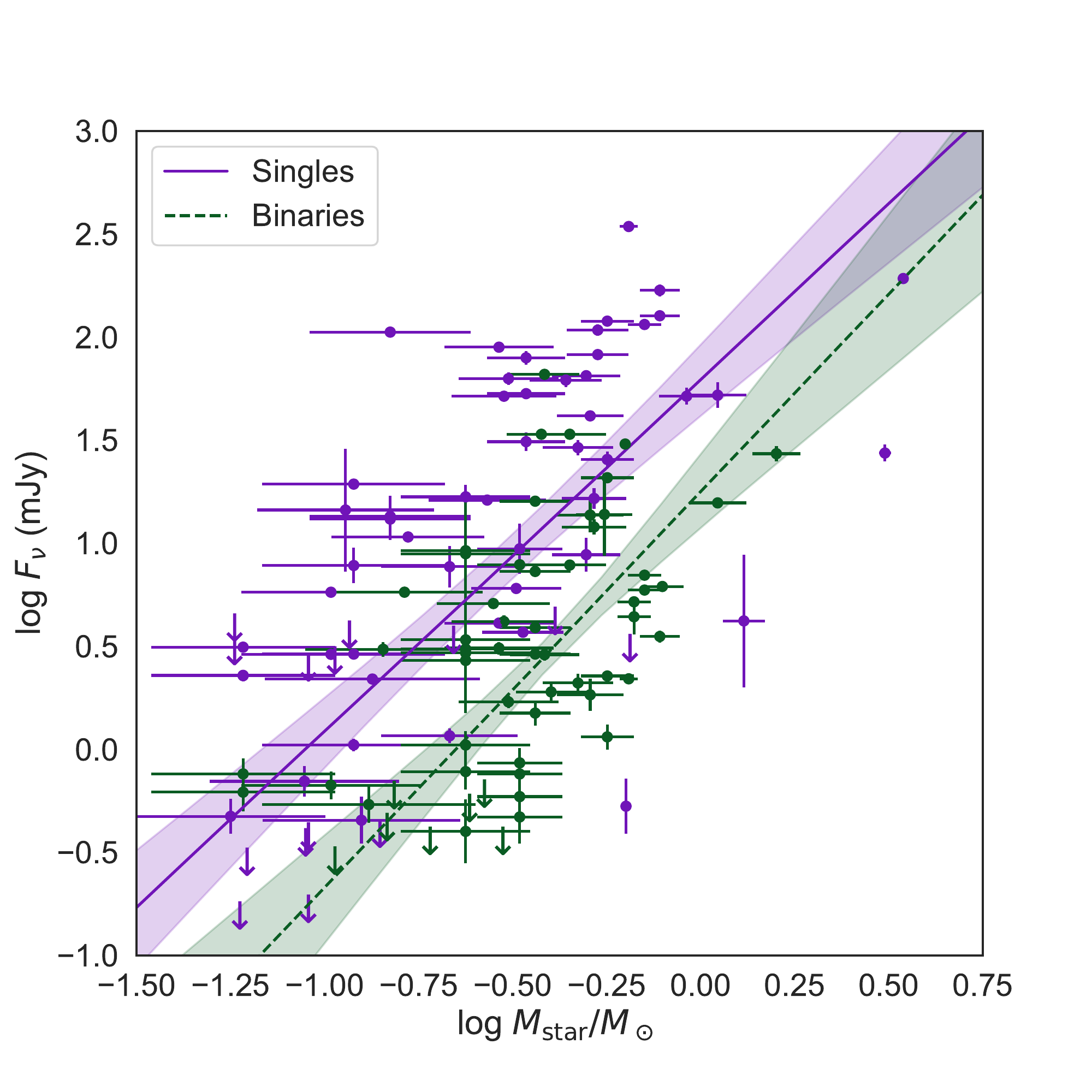}
    \caption{The linear regression fits to the separate samples of single stars (purple) and stars in binary systems (green). The shaded areas show the 68\% confidence regions around the fits.}
    \label{fig:linear}
\end{figure}

As previously shown in several studies in Taurus \citep{har12,and13,pas16} the millimeter flux and derived disk mass depend on the stellar mass.  We see the same impact of stellar mass with the addition of our ALMA detections and more stringent disk mass limits at lower stellar masses (Figure \ref{fig:flux}).  In applying the two sample test to the 1.3 mm flux distribution of single, primary, and secondary sources, we see the same results as \citet{har12}; the distributions of single and primary star fluxes are similar ($p = 0.95$), while the secondary fluxes arise from a different sample than
the singles ($p=0.027$) or the primaries ($p=8.3 \times 10^{-3}$) (Table \ref{tab:twosample}).  
We also fit a linear regression to the millimeter flux as a function of stellar mass separately for our single star and binary samples (Figure \ref{fig:linear}).  We followed the same procedure as \citet{and13}, using the {\it LINMIX\_ERR} MCMC algorithm of \citet{Kelly2007}, as implemented in the Python package {\it linmix\/} by Josh Myers.  The slopes are consistent with each other and broadly consistent with those found by \citet{and13} and \citet{pas16} (1.73$^{+0.27}_{-0.27}$ for single stars and 1.93$^{+0.42}_{-0.39}$ for binary stars), but the intercepts of the two samples are clearly different ($1.80^{+0.17}_{-0.18}$ for the single stars and $1.26^{+0.17}_{-0.18}$ for the individual stars in binary systems), with a lower millimeter flux for the stars in binaries at a given stellar mass. 

\begin{deluxetable}{lllll}
\tablewidth{0pt}
\tablecaption{Two-Sample Comparisons \label{tab:twosample}
}
\tablehead{
\colhead{Sample}    & \multicolumn{4}{c}{p-value} \\
& \colhead{M$_{\ast}$} & \colhead{F$_{\nu}$}    & \colhead{M$_{disk}/M_{\ast}$} 
& \colhead{M$_{disk}/M_{\ast}^{1.3}$} 
}
\tablenum{6}
\startdata
Singles/Primaries & $3.5\times10^{-5}$ & 0.95  & 0.015 & 0.0089\\ 
Singles/Secondaries & 0.26 & 0.027  & $3.0\times10^{-3}$ & 0.0037\\
Primaries/Secondaries & $6.3\times10^{-5}$ & $8.3\times10^{-3}$  & 0.60 & 0.98\\ \hline
Singles/Wide & 0.0050 & 0.93  & 0.18 & 0.11\\
Singles/Close & 0.037 & 0.032  & $1.8\times10^{-4}$ & $2.3\times10^{-4}$\\
Wide/Close & 0.072 & 0.034  & 0.044 & 0.050 \\
\enddata
\end{deluxetable}

\begin{figure}
\includegraphics[width=\textwidth]{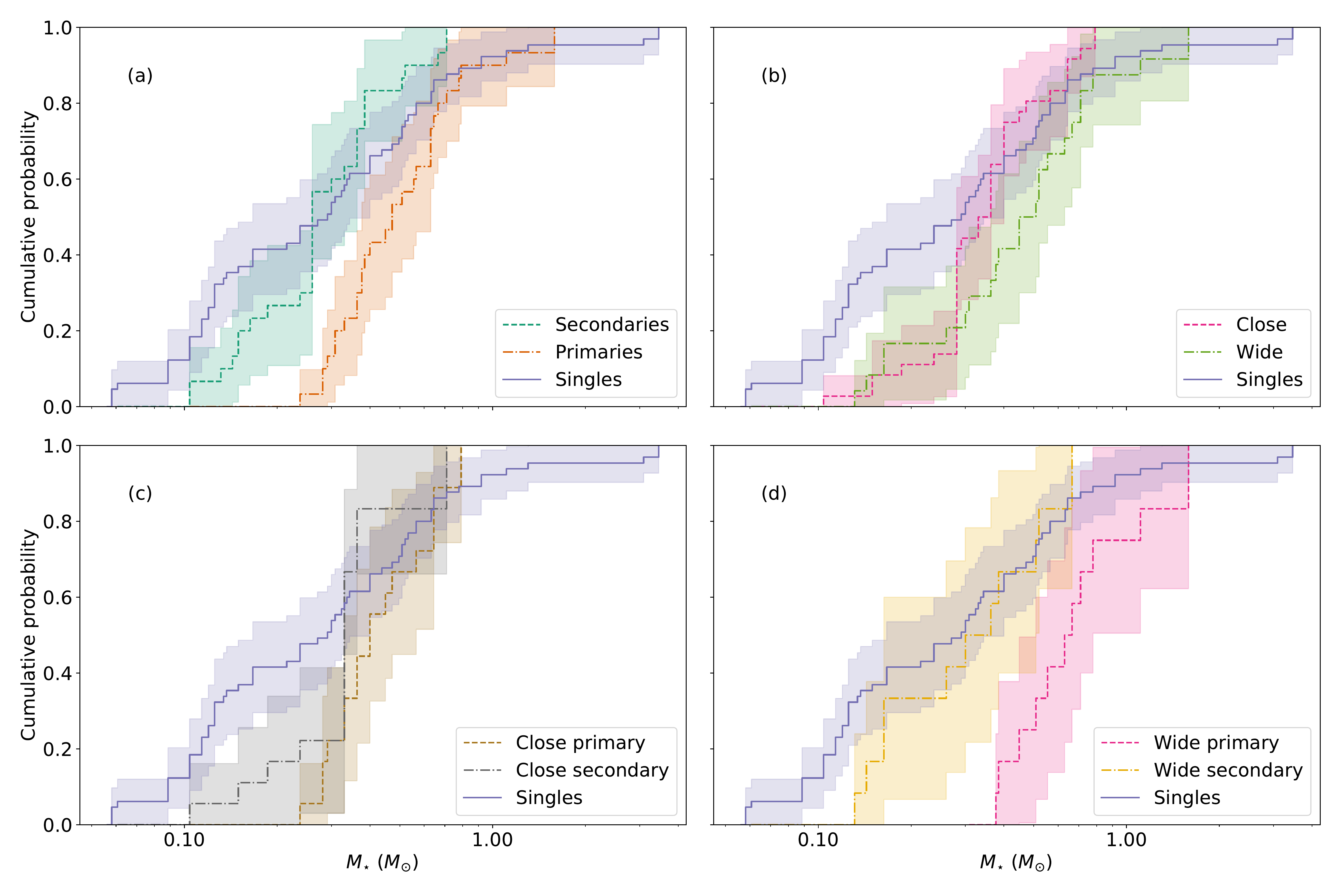}

\caption{Kaplan-Meier estimator of the cumulative distribution of stellar masses for our statistical analysis sample, displayed in various ways.  In this and the following plots, the shaded region shows the $\pm 1\sigma$ confidence region around the estimated distribution. The difference between primary and secondary star stellar masses is more prominent among the wider binaries.}
    \label{fig:stellarmass}
\end{figure}

However, the stellar mass distributions of primary, secondary, and single stars are different from each other, which in turn influences their disk properties.  Thus, to fully understand the comparison of single and binary stars, it is important to take into account the underlying stellar populations.  For the sample
as defined here, including the stellar mass cutoff at M6, the stellar masses for the singles and primaries are drawn from different samples ($p=3.5 \times 10^{-5}$, Table \ref{tab:twosample}).  This is easily seen in plots of the KM estimator of the cumulative distribution; the single stars span the complete mass range, but the distribution of primary stars has a lower mass cutoff at 0.25 M$_{\odot}$ (Figure \ref{fig:stellarmass}).  Unsurprisingly, the distribution of primary stellar masses is also distinct from (and on average larger than) that of secondary stellar masses.  There are some differences between the stellar mass distributions between the wide and close binaries, particularly the wide and close secondaries.    This difference may be due to the small number of objects in each of our binary separation categories as 
a much larger study by \citet{kra11} of 90 Taurus binaries showed no correlation between the mass ratio distribution and the separation distribution.

\begin{figure}
\includegraphics[width=\textwidth]{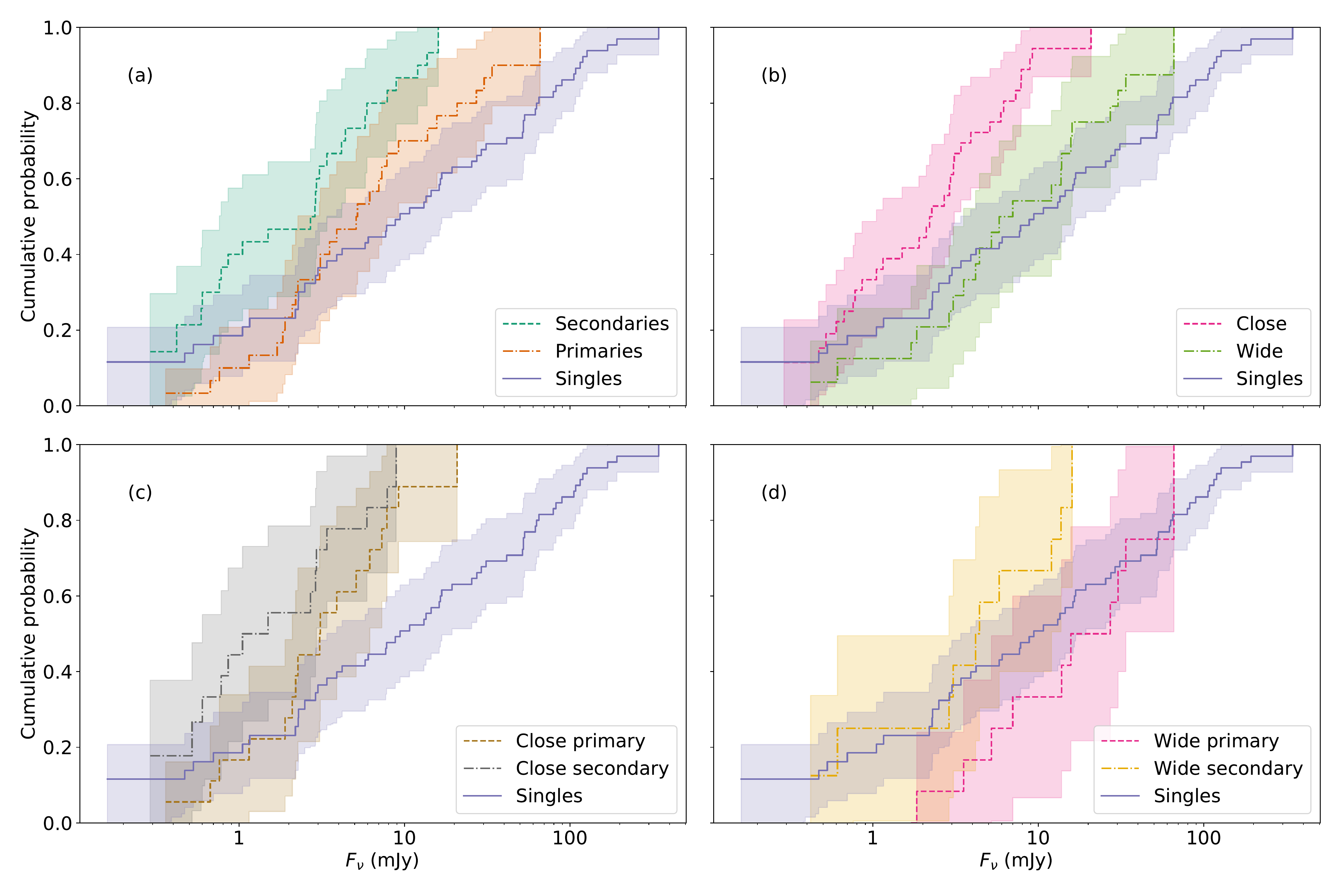}
\caption{Millimeter fluxes for singles and binaries in the statistical sample, divided in various ways as noted.  As seen in previous work, the components of close binaries have significantly lower millimeter fluxes than either single stars or the components of wide binaries.}
    \label{fig:mmflux}
\end{figure}

While the millimeter flux distribution plots (Figure \ref{fig:mmflux}) match previous work \citep{har12,and13}, we also looked at other metrics to further investigate the role of stellar mass.
Fitting a linear regression to the disk mass as a function of the stellar mass, using the same procedure as for flux above, yields slopes and intercepts of $1.26^{+0.26}_{-0.26}$ and $-2.32 \pm 0.17$ for the single stars and 
$1.52^{+0.41}_{-0.40}$ and $-2.85\pm{0.18}$ for the binary stars.  These slopes are consistent with \citet{and13} and while the slopes are consistent within 1 $\sigma$, the confidence intervals around the best-fit lines are clearly distinct from each other (similar to what is shown for the fluxes in Figure 7), with the singles having larger typical disk masses at a given stellar mass.  We compared the samples of single, primary, and secondary stars for both  $\log \left( M_{\rm disk} / M_{\star} \right)$ and $\log \left( M_{\rm disk} / M_{\star}^{1.3}\right) $ with the results given in Table \ref{tab:twosample}.  For both metrics, the single stars are drawn from a different population than either the primaries or the secondaries, while the primary and secondary stars are consistent with being drawn from the same population.
As the statistics are similar between these two metrics, we have used $\log \left( M_{\rm disk} / M_{\star} \right)$ in our plots to facilitate comparison with works examining other star formation regions such as Chamaeleon \citep{pas16}.  We note that our observed similarity of $\log \left( M_{\rm disk} /  M_{\star}\right)$  for primary and secondary stars is a different result than found in previous work \citep[e.g.][]{har12}.  The difference may result from our much higher detection rate among low-mass stars, both among the single stars and the secondaries.

The previously reported tendency for the disks around primary stars to be brighter than disks around secondaries at millimeter wavelengths is largely an artifact of their greater stellar masses; when the different stellar masses of primaries and secondaries are accounted for, there is no clear difference in the distribution of their disk properties.

\subsection{The influence of binary separation}

\begin{figure}
\includegraphics[width=\textwidth]{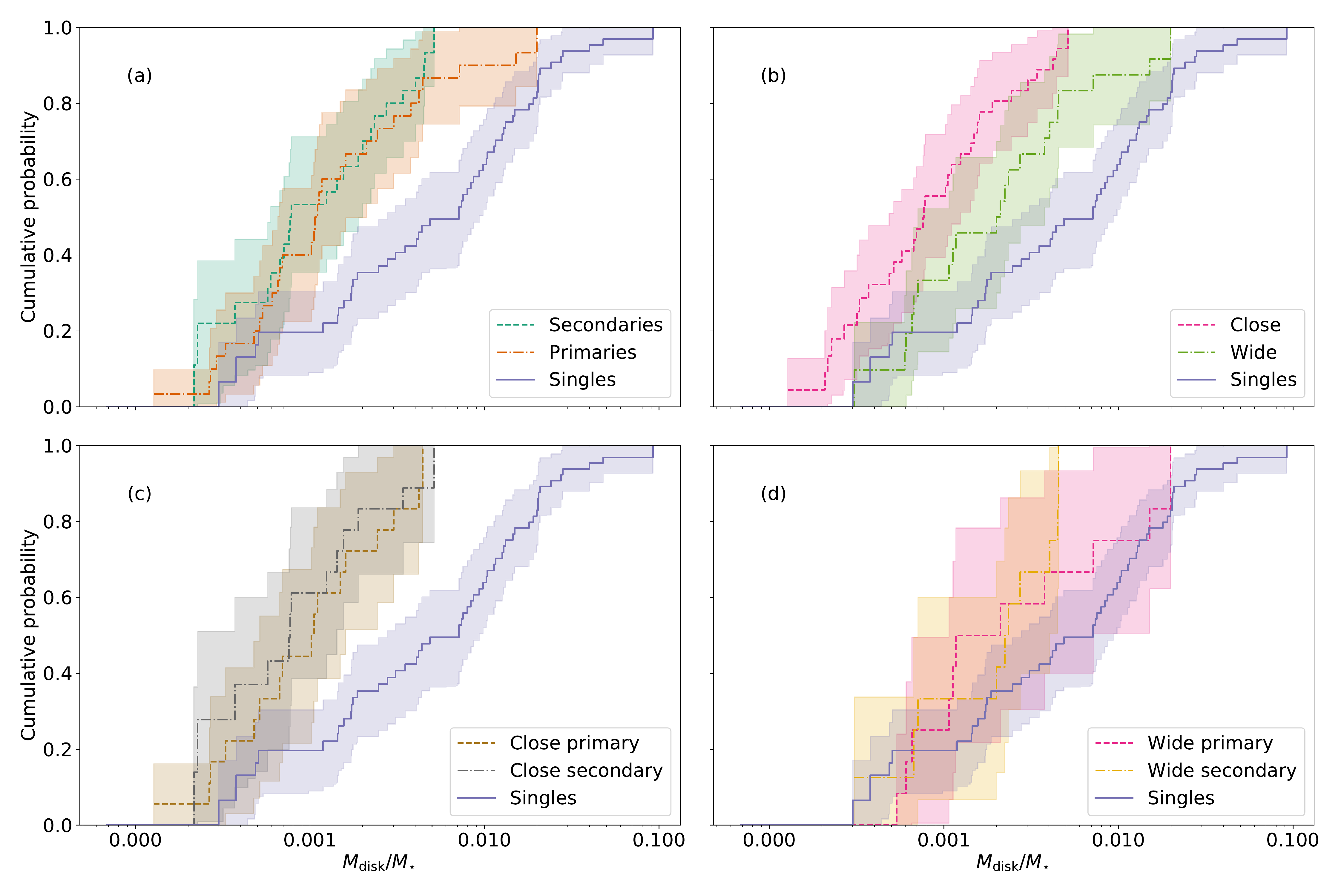}
\caption{Disk/stellar mass ratio.  Note that primary and secondary stars have indistinguishable distributions of disk mass once we normalize by the stellar mass (panel a), and that both are clearly lower than single star disks.  In addition, the closer binaries as a group not only have lower millimeter fluxes than single stars (as seen in the previous figure), but also lower disk to star mass ratios.  Comparing panel b in this and the previous figure, scaling by stellar mass also reveals more of a difference between singles and wide binaries in the disk/star mass ratio than seen in the flux alone, indicating that even the wider binary disks may have been influenced by multiplicity. }
    \label{fig:disktostar}
\end{figure}

\begin{figure}
    \centering
    \includegraphics[angle=270,width=5.5in]{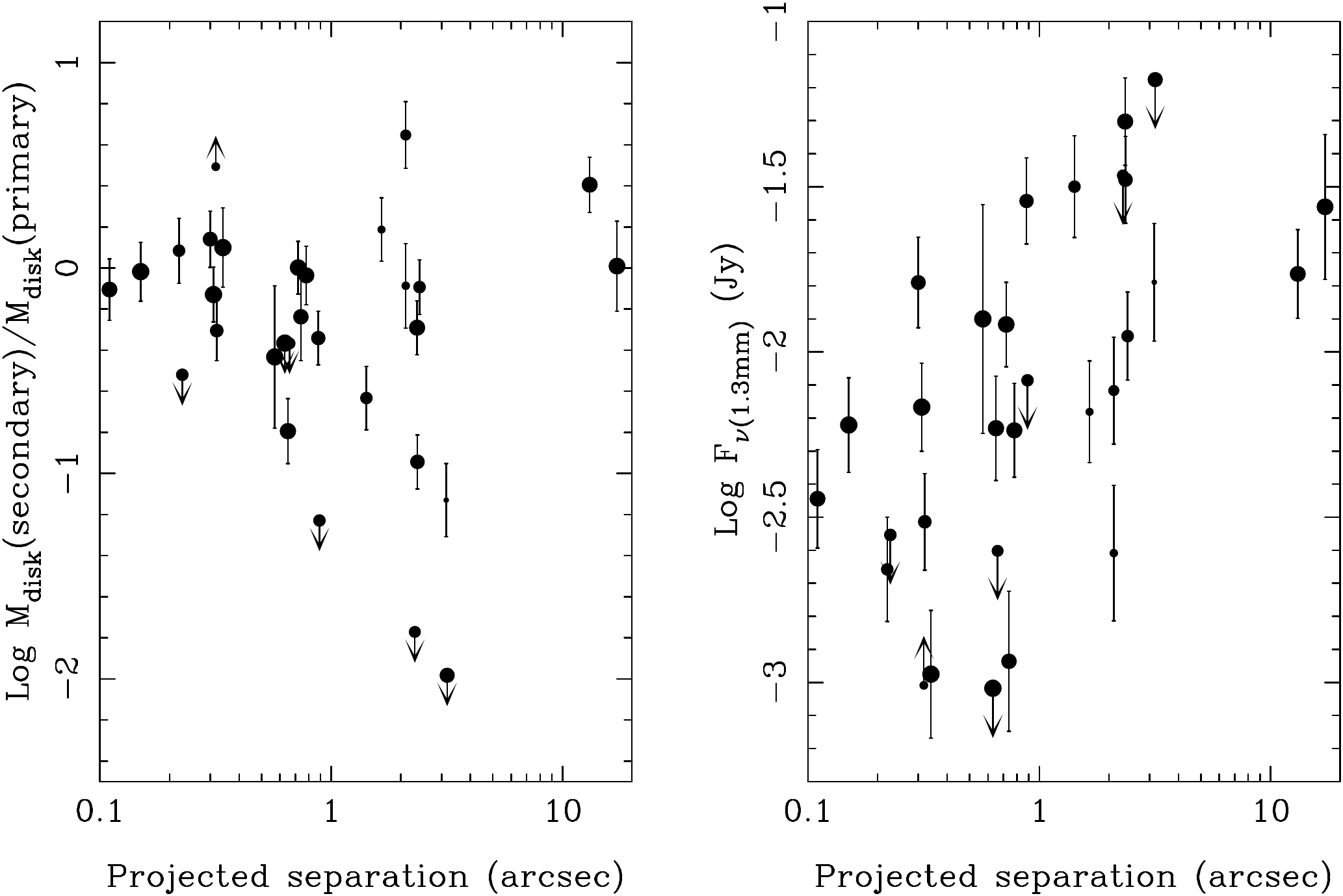}
    \caption{Left: The disk mass ratio (secondary/primary) plotted against projected separation for the binary star sample as defined in \S \ref{sec:binary}.  The symbol size is scaled to the stellar mass ratio, with larger points for binaries with equal mass ratios. Right: The total 1.3 mm flux versus the projected separation for the binary sample.}
    \label{fig:separation}
\end{figure}

Both the two-sample comparison (Table \ref{tab:twosample}) and the KM plots (Figures \ref{fig:mmflux} and \ref{fig:disktostar}) show that the millimeter flux and disk mass of wide ($>140$ AU) binaries are much closer to those of single stars than close binaries. This impact on the disk properties of the binary separation has been seen previously \citep{jen96,har12}. Our more sensitive survey finds that many close binary systems do have disks, but with lower flux than their wider separation counterparts.   
In Figure \ref{fig:separation} we plot the ratio of the secondary to primary disk mass and the total (primary + secondary) 1.3 mm flux as a function of projected separation.  In this sample, close binaries tend to have roughly equal mass disks, while wide binaries have a much wider range of disk mass ratios.    In both plots, the point size is scaled to the stellar mass ratio, with larger points corresponding to more equal stellar mass ratios.  There is no obvious correlation with stellar mass ratio in either of these plots.  

One unresolved question with the smaller flux from binary disks is whether this is due to intrinsically smaller disks, perhaps due to truncation, or to different disk structure.  Recent studies by \citet{tri17} and \cite{and18} have used resolved disk samples to constrain the contribution of optically thick emission and have confirmed the disk size/luminosity relation previously inferred from smaller samples.   These samples were specifically constructed to exclude close binary stars and it would be instructive to compare the disk size between our single and binary samples.  However, while our ALMA data for the close binaries has sufficient angular resolution to resolve or constrain the disk size, this is not the case for the majority of the wide binaries and single stars.  For instance, of the 66 stars in our singles sample, only 16 have sufficient data to be included in the \citet{and18} study.  Additional high resolution observations of the complete single and wide binary sample would allow a comparison of disk size with the same categories as our flux comparisons.

In the plot of total system flux as a function of separation (Figure \ref{fig:separation}), there is a clear trend of higher total flux for larger separations;   in our sample, the median flux for the close (15 to 140 AU)  binaries is 4.7 mJy, while the median flux for the 140 to 4200 AU binaries is 22.4 mJy.  \citet{har12} used different separation divides of $<$30 AU, 30--300 AU and $>$300 AU and included pairs from higher order multiples.
Given our sample definition, we only have two systems in their lowest bin, but using the same definitions for the two larger bins, we see roughly the same factor of 5 increase in flux.

\subsection{Comparison of primary and secondary disks}

In comparing the properties of the primary and secondary circumstellar disks, we note several properties from this sample:
\begin{enumerate}
    \item The two-sample tests indicate that the $M_{\rm disk}/M_{\star}$ values for primary and secondary stars are consistent with being drawn from the same distribution ($p=0.60$).  As noted in \S \ref{sec:stellarmass}, both of these samples are drawn from a different population than the single stars for this parameter.  The KM plots show that $M_{\rm disk}/M_{\star}$ is systematically higher for the single stars (Figure \ref{fig:disktostar}).  Dividing the binaries into wide and close samples instead shows that the wide binaries are closer to the single distribution, but are still systematically lower.  Previous studies have shown this negative impact on disk mass for multiple systems \citep{jen96, har12, cox17} but here we show that {\it as a function of  stellar mass}, both the primary and secondary millimeter fluxes and disk masses are systematically lower than for single stars (see also Figure \ref{fig:linear}).

    \item It has been shown in previous large studies of Taurus and other star formation regions that while the disk mass is generally correlated with stellar mass, there is roughly an order of magnitude scatter in disk mass around this relationship \citep{and13,pas16,ans16,bar16}.  If this scatter is due to physical quantities or processes that vary across the star formation region but are constant on scales of stellar binaries, we would expect the disk mass ratio and the stellar mass ratio in binary systems to be correlated. The left panel of Figure \ref{fig:primarysecondary} plots this relation for our binary sample, with the diagonal line tracing equal ratios.  Consistent with Paper 1, these binary systems show no correlation between disk mass ratio and stellar mass ratio, for either wide or close systems. Using the Kendall $\tau$ correlation, the p-value for no correlation is 1 for the wide systems and 0.3 for the close. This lack of correlation can be explained in two ways: 1) The parameters that cause the scatter in the general disk-stellar mass relation vary on scales of less than 30 AU or 2) the processes controlling the disk mass in binary formation are different than those for single stars and apply even to binaries with separations much larger than the canonical disk size.  We note that the scatter is both above and below the equality line; i.e.\ for some systems the primary disk is more massive than expected from the stellar mass ratio, while in others the secondary disk is more massive than expected.
    
\begin{figure}
    \centering
    \includegraphics[angle=270,width=6.5in]{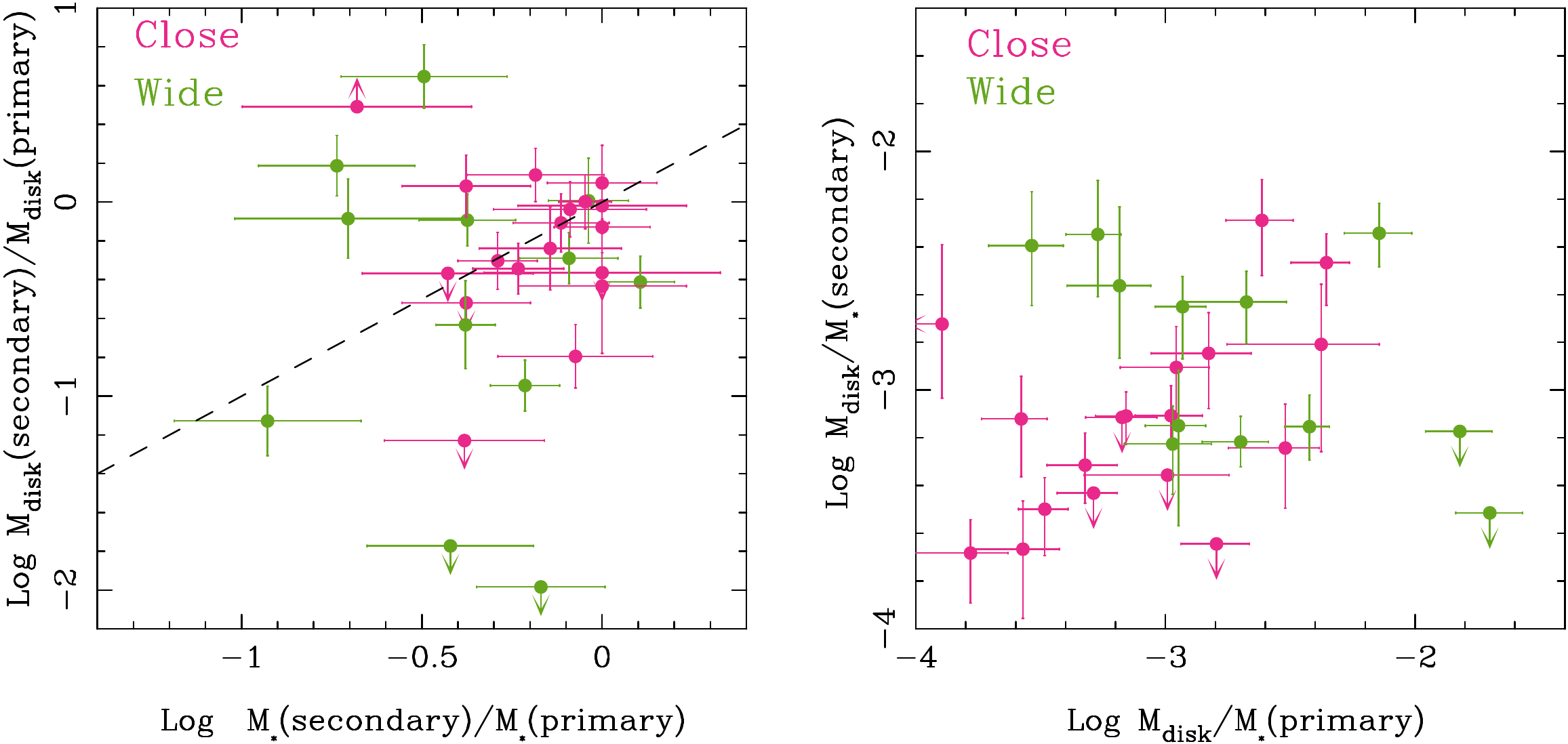}
    \caption{Left: the disk mass ratio for each binary system (secondary/primary) plotted against the stellar mass ratio. Close binaries ($<100$ AU) are plotted in pink, while wide binaries are plotted in green.  The dashed line represents the equal ratios between the x and y axes and is not a fit to the data. Right: Log $M_{\rm disk}/M_{\star}$ for the secondary star plotted against the same ratio for the primary star.  }
    \label{fig:primarysecondary}
\end{figure}    

\item  The sensitivity and angular resolution of our ALMA observations have resulted in a much higher detection rate for secondary components than found by previous surveys in Taurus \citetext{\citealt{har12}; Paper 1}.  In these Class II binaries, 
the primary disk mass does not dominate the total disk mass in the binary system (Figure \ref{fig:separation}).  The average fractional disk mass for the primary compared to the total disk mass in the binary is 62\%, with our assumption of optically thin emission. While there are binaries where M$_{\rm disk,\, secondary} < 0.1 $M$_{\rm disk,\, primary}$, there are also three systems where the ratio of the secondary to primary disk mass is significantly ($>2\sigma$) higher than one.  These systems span the range of projected separations but in each case, the spectral types are at least three sub-classes apart, so the primary and secondary stellar masses are significantly different.  These systems contradict the predictions of binary star formation by \citet{bat00} that in general, the primary disk will have more mass than the secondary disk; however, these models do not include disk evolution.  The three systems are:
\begin{itemize}
    \item CZ Tau:  Only the secondary disk is detected for CZ Tau and the secondary/primary disk mass limit is $>3.1$.   The position of the source we detect with ALMA has an uncertainty of less than 20 mas (Table \ref{tab:obs}) and is located at a separation of 0\farcs397 and position angle of 87.5\arcdeg\ from the source detected in Gaia DR2 \citep{gaia18}, to be compared with the separation of 0\farcs317 and position angle of $88.1\arcdeg \pm 1.1\arcdeg$ reported by \citet{whi01}.   If the single Gaia detection is the primary, then the position of the ALMA source agrees well with the expected position of the secondary based on past observations, with a modest separation change due to orbital motion as we see for most other systems in this separation range (Table \ref{tab:BinarySep}).  If Gaia were seeing the secondary, then the source we see with ALMA would be 180\arcdeg\ from where the primary should be but coincidentally at the same separation. The spectral types are M3 and M6.
    \item CoKu Tau 3: Both components of this binary were detected in Paper 1 and the spectral types are M0.5 and M4.3.  The secondary/primary disk mass ratio is $4.4\pm1.6$.
    \item GI/GK Tau: These disks were both detected in previous surveys.  GK Tau is the primary with a spectral type of K6.5 while GI Tau has a spectral type of M0.4.  The secondary/primary disk mass ratio is $2.5\pm0.8$.
\end{itemize}

\item  A different probe of the relationship between the primary and secondary disks is to compare $M_{\rm disk}/M_{\star}$ within the binary system; i.e. if the primary has a low or high value of $M_{\rm disk}/M_{\star}$, is that also seen for the secondary?  This comparison is shown in the right panel of Figure \ref{fig:primarysecondary}.  As can be seen, the wide and close binaries follow different patterns.  To confirm this, we used the Kendall's $\tau$ relation to test the correlation for the two samples, and the Akritas-Theil-Sen method to find the estimated slope, again using the R routine {\it cenken}.  The slopes and  correlation coefficients $\tau$  are given in Table \ref{tab:cenken}, where $\tau=0$ corresponds to no correlation and the $p$-value is reported for no correlation being present (i.e.\ a low $p$-value value represents the presence of a correlation).  For all binaries, there is no correlation between the primary and secondary ratios ($\tau = 0.08, p=0.53$).  Analyzing the wide and close samples separately, the $\tau$ and $p$-values indicate there is a correlation. However, while the close binaries are positively correlated, the wide binaries are negatively correlated.
That is, in wide binary systems, if the primary has a high 
value of $M_{\rm disk}/M_{\star}$, the secondary preferentially has a low value of $M_{\rm disk}/M_{\star}$.
One possible explanation is that if close binaries have disks that accrete from a common envelope, whatever mechanism controls the division of mass between the disk and the star operates similarly for both stars.  On the other hand, for the wider binaries, the mass accretion process may be competitive, and if the primary disk receives more mass relative to the star, this could result in less relative disk mass for the secondary.

\end{enumerate}

\begin{deluxetable}{llll}
\tabletypesize{\scriptsize}
\tablewidth{0pt}
\tablecaption{Correlation tests of Primary vs.\ Secondary Disk/Stellar Mass Ratios\label{tab:cenken}
}
\tablehead{
\colhead{Sample} & \colhead{Slope\tablenotemark{a}}   & \colhead{Kendall's $\tau$} & \colhead{$p$\tablenotemark{b}} 
}
\tablenum{7}
\startdata
All & 0.22 & 0.08 & 0.53 \\
Wide & $-$0.48 & $-$0.35 & 0.12 \\
Close & 0.62 & 0.37 & 0.031 \\
\enddata
\tablenotetext{a}{Nonparametric slope estimate from Akritas-Theil-Sen method}
\tablenotetext{b}{Probability that no correlation is present.}
\end{deluxetable}

\section{Conclusion}

We have conducted a survey of Class II sources in Taurus with ALMA and made 21 new detections.  Combining our data with that of previous surveys, the detection fraction for Class II stars in Taurus is now above 70\% for stars in both single and multiple systems.  This high detection fraction makes this sample ideal for comparison of disk properties, particularly the impact of multiplicity.  We find the following properties for disks around Class II stars in Taurus:

\begin{itemize}
    
    \item Comparing the observed flux as a function of stellar mass for single, primary, and secondary disks, the single stars clearly have 
    a higher flux level, while primary and secondary stars have the same flux levels for a given spectral type.   Previous studies that showed similar flux levels for single and primary stars for a similar range of binary separations did not control for the difference in stellar mass distributions between the single and primary star populations.  Similarly, in comparisons of $M_{\rm disk}/M_{\star}$, single stars are drawn from a different population than both primary and secondary stars, while primary and secondary stars appear to be drawn from the same population. 
    
    \item The disk mass from the primary star accounts for 62\% of the total mass on average and thus does not dominate the total disk mass for these Class II binary systems. The range of disk mass ratios spans almost three orders of magnitude, although the stellar mass ratios only span one order of magnitude. For three binaries in our sample, the circumsecondary disk mass is higher than the circumprimary disk mass and this occurs across a range of projected separations.
    
    \item Most close ($< 140$ AU) binaries do have circumstellar disks around one or both components, but these disks are significantly less massive than around wider binaries or single stars.  This applies even to the closest binaries in our sample (15 to 60 AU) where the detection fraction is 87\% for both primaries and secondaries.  We detect no circumbinary disks; if there is any circumbinary material in these systems, it must be extremely tenuous to have escaped detection in this survey. 
    
    \item In close ($< 140$ AU) binaries the values of $M_{\rm disk}/M_{\star}$ for the primary and secondary stars are correlated, but in wide binaries, they are anti-correlated.
    
\end{itemize}

This paper makes use of the following ALMA data:
ADS/JAO.ALMA\#2013.1.00105.S and \\ ADS/JAO.ALMA\#2015.1.00392.S.
ALMA is a partnership of ESO (representing its member states), 
NSF (USA) and NINS (Japan), together with NRC (Canada), NSC and ASIAA (Taiwan), and KASI 
(Republic of Korea), in cooperation with the Republic of Chile. 
The Joint ALMA Observatory is operated by ESO, AUI/NRAO and NAOJ.
The National Radio Astronomy Observatory is a facility of the National Science Foundation operated under cooperative agreement by Associated Universities, Inc.

JMC acknowledges support from the National Aeronautics and Space Administration under Grant No. \\15XRP15\_20140 issued through the Exoplanets Research Program.  We thank Adam Kraus, Lisa Prato, and Gail Schaefer for helpful discussions, and Kim Ward-Duong and Jenny Patience for sharing their results before publication.  

This research made use of Astropy, a community-developed core Python package for Astronomy \citep{Astropy2018};   matplotlib, a Python library for publication quality graphics \citep{Hunter2007}; NASA's Astrophysics Data System; and the SIMBAD database, operated at CDS, Strasbourg, France.

\facilities{ALMA}
\software{CASA \citep{mcm07}, R, Astropy \citep{Astropy2018}, matplotlib \citep{Hunter2007}}

\section*{Appendix A: Notes on CO emission from individual sources}

Peak fluxes and RMS values from the integrated CO emission are presented in Table \ref{tab:CO}.  Here we briefly discuss the sources detected in CO.

\begin{itemize}
\item[] 
{\it 2MASS J04295950+2433078:}  Weak CO detection. 

\item[]
{\it CIDA 9 (J05052286+2531312):}  This source is strongly detected and spatially resolved in the continuum map and the CO emission shows a clear detection and east-west velocity trend (Figure \ref{fig:cida9}). 

\item[]
{\it DP Tau:}  The CO emission is extended along the binary position angle, similar to what is seen in the continuum (Figure \ref{fig:multis}), suggesting that there is CO associated with both components.

\item[]
{\it FX Tau}:  The primary source, detected in the continuum, is also detected in CO\null.  The CO emission centroid is offset by $\sim 0\farcs08$ from the continuum centroid; this appears to be the emission from one side of the disk, with the other side's emission falling at velocities that are similar to those where absorption from the molecular cloud is seen in other sources in Taurus, and thus presumably absorbed by the cloud. 

\item[]
{\it GG Tau:}  CO emission associated with both the central source and the large circumbinary ring is clearly detected, though our sensitivity to this low surface brightness emission is not nearly as good as previously published ALMA CO data; see \cite{Phuong2018}.

\item[]
{\it GH Tau:} The western source is weakly detected in CO.

\item[]
{\it Haro 6-28:}  There are weak CO detections at the positions of both
binary components. As with FX Tau, it appears that only half of the primary's disk is detected, with the other half at velocities that are absorbed by the cloud.

\item[]
{\it Haro 6-37:}  We detect component C (NE at 2\farcs7) in CO, as in the continuum.  CO emission is not apparent until velocities $\gtrsim 8$ km s$^{-1}$, suggesting that half of the disk emission may be absorbed by the cloud, as with some other sources.  There is some complicated emission structure at higher velocities.  The position angle of the emission is roughly $20\arcdeg$. 

\item[]
{\it HN Tau:}  Very compact CO detection with a possible velocity gradient at PA $\sim 135\arcdeg$. 

\item[]
{\it ITG 33A:}  The source detected in the continuum map is also weakly detected in CO. 

\item[]
{\it IRAS 04187+1927:}  Modest CO detection with a possible velocity gradient at PA of roughly 100\arcdeg.

\item[]
{\it UX Tau:}  The CO emission from the ring (Figure \ref{fig:uxtau}) shows a clear rotational signature, with the northern edge receding. 

\item[]
{\it V807 Tau:}  There is a clear CO detection of the primary, though many channels are affected by cloud absorption. 

\item[]
{\it V710 Tau:}  Only the northern source is detected in the continuum, and CO emission is present at that position as well, with a velocity gradient at a position angle of roughly 250\arcdeg. 

\item[]
{\it VY Tau:}  The primary source detected in the continuum map is also marginally detected in CO. 

\item[]
{\it XZ Tau:}  Complicated emission pattern; both sources seen in the continuum map appear to be detected, but there is additional structure present in the CO emission as well. 

\end{itemize}

\begin{deluxetable}{lll}
\tabletypesize{\scriptsize}
\tablewidth{0pt}
\tablecaption{CO detections \label{tab:CO}
}
\tablehead{
\colhead{Source}    & \colhead{Peak flux}   & \colhead{RMS} \\
 & \multicolumn{2}{c}{(Jy beam$^{-1}$ km s$^{-1}$)}
}
\tablenum{8}
\startdata
CIDA 9 (J05052286+2531312) & 0.284 & 0.049 \\
DP Tau & 0.323 & 0.066 \\
FX Tau & 0.292 & 0.043 \\
GG Tau & 0.295 & 0.043 \\	
GH Tau & 0.159 & 0.038 \\
Haro 6-28 & 0.134 / 0.152 & 0.041 \\
Haro 6-37 & 0.253 & 0.043 \\
HN Tau & 0.749 & 0.058 \\
IRAS 04187+1927 & 0.452 & 0.058 \\
ITG 33A & 0.192 & 0.065 \\
J04295950+2433078 & 0.267 & 0.060 \\
UX Tau	& 0.467 & 0.039	\\
V807 Tau & 0.343 & 0.039 \\
V710 Tau & 0.810 & 0.058 \\
VY Tau	& 0.239 & 0.076 \\
XZ Tau	& 0.631 / 0.469 & 0.048 \\
\enddata
\end{deluxetable}

\begin{longrotatetable}
\begin{deluxetable}{llrlllllrrrrl}
\tabletypesize{\scriptsize}
\tablewidth{0pt}
\tablecaption{Taurus sample disk and stellar properties \label{tab:sample}
}
\tablehead{
\colhead{2MASS name} & \colhead{Name}  & \colhead{Role\tablenotemark{a}} &\colhead{Samp.} & \colhead{Binary}  & \colhead{Flux} & \colhead{Ref\tablenotemark{b}} & \colhead{Sp. type} & \colhead{Ref\tablenotemark{b}} & \colhead{log M$_{\ast}$} &  \colhead{$\log L_{\ast}$} & \colhead{log M$_{\rm disk}$} & \colhead{Distance} \\
& & & & sep. (\arcsec)\tablenotemark{b} & \colhead{(mJy)} & & & & \colhead{(M$_{\odot}$)} & \colhead{(L$_{\odot}$)} & \colhead{(M$_{\odot}$)}  & \colhead{(pc)\tablenotemark{c}}
}
\tablenum{4}
\startdata
J04135737+2918193 & IRAS 04108+2910 & -9  & N & ... & $<$19.80 & 8 & M3 & 12 & $-0.54_{-0.19}^{+0.10}$ & $-0.50_{-0.19}^{+0.12}$ & $<$-2.73 & 123.07$\pm$1.53 \\ 
J04141188+2811535 & J04141188+2811535 & 0  & N & ... & 0.28$\pm$0.08 & 11 & M6.25 & 13 & $-1.25_{-0.20}^{+0.27}$ & $-1.57_{-0.32}^{+0.43}$ & -4.25$\pm$0.16 & 130.70$\pm$2.86 \\ 
J04141358+2812492 & FM Tau  & 0  & Y & ... & 13.10$\pm$2.70 & 8 & M4.5 & 14 & $-0.83_{-0.23}^{+0.20}$ & $-0.85_{-0.39}^{+0.26}$ & -2.76$\pm$0.13 & 131.44$\pm$0.81 \\ 
J04141458+2827580 & FN Tau & 0  & Y & ... & 16.80$\pm$2.10 & 8 & M3.5 & 14 & $-0.62_{-0.20}^{+0.14}$ & $-0.59_{-0.26}^{+0.16}$ & -2.72$\pm$0.10 & 130.76$\pm$1.06 \\ 
J04141700+2810578 & CW Tau  & 0  & Y & ... & 52.30$\pm$7.00 & 8 & K3.0 & 14 & $0.05_{-0.08}^{+0.07}$ & $0.38_{-0.14}^{+0.11}$ & -2.46$\pm$0.10 & 131.94$\pm$0.68 \\ 
J04141760+2806096 & CIDA 1 & 0  & Y & ... & 13.50$\pm$2.80 & 8 & M4.5 & 14 & $-0.83_{-0.23}^{+0.20}$ & $-0.85_{-0.39}^{+0.26}$ & -2.72$\pm$0.13 & 135.19$\pm$1.59 \\ 
J04142626+2806032 & MHO 1 & 1  & N & ... & 216.00$\pm$0.76 & 9 & M2.5 & 15 & $-0.48_{-0.14}^{+0.08}$ & $-0.43_{-0.16}^{+0.11}$ & -1.64$\pm$0.09 & 132.32$\pm$3.85 \\ 
J04142626+2806032 & MHO 2 AB & -2  & N & ... & 133.30$\pm$0.79 & 9 & M2.5/5.2 & 15 & $-0.18_{-0.14}^{+0.08}$ & $-0.35_{-0.20}^{+0.17}$ & -1.87$\pm$0.09 & 132.32$\pm$3.85$^{1}$ \\ 
J04143054+2805147 & BHS98 MHO 3 AB & -1  & N & ... & $<$12.00 & 8 & K7/M2 & 15/8 & $0.11_{-0.01}^{+0.04}$ & $0.14_{-0.04}^{+0.09}$ & $<$-2.52 & 241.36$\pm$11.29 \\ 
J04144730+2646264 & FP Tau & 0  & Y & ... & 6.04$\pm$0.20 & 11 & M2.6 & 14 & $-0.49_{-0.16}^{+0.08}$ & $-0.44_{-0.17}^{+0.11}$ & -3.22$\pm$0.09 & 128.01$\pm$0.85 \\ 
J04144786+2648110 & CX Tau  & 0  & Y & ... & 9.40$\pm$2.30 & 8 & M2.5 & 14 & $-0.48_{-0.14}^{+0.08}$ & $-0.43_{-0.16}^{+0.11}$ & -3.04$\pm$0.12 & 127.46$\pm$0.65 \\ 
J04144928+2812305 & FO Tau A & 1  & Y &  0.150$^{1}$ & 3.07$\pm$0.19 & 7 & M3.5 & 16 & $-0.62_{-0.20}^{+0.14}$ & $-0.59_{-0.26}^{+0.16}$ & -3.45$\pm$0.09 & 132.07$\pm$1.50$^{2}$ \\ 
J04144928+2812305 & FO Tau B & 2  & Y &  0.150$^{1}$ & 2.94$\pm$0.19 & 7 & M3.5 & 16 & $-0.62_{-0.20}^{+0.14}$ & $-0.59_{-0.26}^{+0.16}$ & -3.47$\pm$0.09 & 132.07$\pm$1.50$^{2}$ \\ 
J04153916+2818586 & J04153916+2818586 & -9  & N & ... & 13.40$\pm$1.40 & 8 & M4.0 & 14 & $-0.73_{-0.19}^{+0.19}$ & $-0.68_{-0.35}^{+0.19}$ & -2.80$\pm$0.10 & 131.01$\pm$1.40 \\ 
J04154278+2909597 & IRAS 04125+2902 & -9  & N & ... & 19.90$\pm$2.50 & 8 & M1.25 & 12 & $-0.37_{-0.09}^{+0.10}$ & $-0.28_{-0.12}^{+0.15}$ & -2.55$\pm$0.10 & 159.24$\pm$1.69 \\ 
J04155799+2746175 & J04155799+2746175 & -9  & N & ... & 12.60$\pm$1.40 & 8 & M5.2 & 14 & $-0.97_{-0.27}^{+0.20}$ & $-1.12_{-0.43}^{+0.39}$ & -2.69$\pm$0.13 & 135.19$\pm$1.97 \\ 
J04161210+2756385 & J04161210+2756385 & 0  & Y & ... & 2.19$\pm$0.12 & 11 & M4.75 & 13 & $-0.87_{-0.38}^{+0.19}$ & $-0.94_{-0.65}^{+0.29}$ & -3.48$\pm$0.11 & 137.00$\pm$2.21 \\ 
J04163911+2858491 &  J04163911+2858491 & 0  & Y & ... & $<$2.50 & 8 & M5.5 & 17 & $-1.05_{-0.27}^{+0.23}$ & $-1.25_{-0.45}^{+0.39}$ & $<$-3.22 & 159.24$\pm$1.69$^{2}$ \\ 
J04173372+2820468 & CY Tau  & 0  & Y & ... & 79.40$\pm$5.90 & 8 & M2.3 & 14 & $-0.46_{-0.12}^{+0.09}$ & $-0.41_{-0.14}^{+0.12}$ & -2.11$\pm$0.09 & 128.41$\pm$0.73 \\ 
J04174955+2813318 & KPNO 10 & 0  & Y & ... & 7.80$\pm$1.40 & 8 & M5 & 18 & $-0.92_{-0.29}^{+0.19}$ & $-1.03_{-0.49}^{+0.35}$ & -2.91$\pm$0.13 & 136.89$\pm$2.20 \\ 
J04174965+2829362 & SS94 V410 X-ray 1 & 0  & Y & ... & $<$3.40 & 8 & M3.7 & 14 & $-0.67_{-0.20}^{+0.17}$ & $-0.63_{-0.29}^{+0.18}$ & $<$-3.42 & 128.57$\pm$1.27 \\ 
J04181078+2519574 & V409 Tau & -9  & N & ... & 18.70$\pm$1.40 & 8 & M0.6 & 14 & $-0.30_{-0.10}^{+0.08}$ & $-0.18_{-0.15}^{+0.11}$ & -2.78$\pm$0.09 & 130.87$\pm$0.69 \\ 
J04181710+2828419 & V410 Anon 13 & 0  & Y & ... & 0.47$\pm$0.08 & 11 & M5.75 & 19 & $-1.25_{-0.13}^{+0.38}$ & $-1.58_{-0.20}^{+0.65}$ & -4.07$\pm$0.19 & 124.31$\pm$5.21 \\ 
J04183112+2816290 & DD Tau A & 1  & Y &  0.570$^{1}$ & 9.20$\pm$4.30 & 8 & M3.5 & 16 & $-0.62_{-0.20}^{+0.14}$ & $-0.59_{-0.26}^{+0.16}$ & -3.00$\pm$0.18 & 128.39$\pm$0.96$^{2}$ \\ 
J04183112+2816290 & DD Tau B & 2  & Y &  0.570$^{1}$ & 3.40$\pm$1.90 & 8 & M3.5 & 16 & $-0.62_{-0.20}^{+0.14}$ & $-0.59_{-0.26}^{+0.16}$ & -3.43$\pm$0.21 & 128.39$\pm$0.96$^{2}$ \\ 
J04183158+2816585 & CZ Tau A & 1  & Y &  0.317$^{1}$ & $<$0.36 & 7 & M3 & 20 & $-0.54_{-0.19}^{+0.10}$ & $-0.50_{-0.19}^{+0.12}$ & $<$-4.43 & 128.39$\pm$0.96$^{2}$ \\ 
J04183158+2816585 & CZ Tau B & 2  & Y &  0.317$^{1}$ & 0.62$\pm$0.12 & 7 & M6 & 8 & $-1.22_{-0.20}^{+0.29}$ & $-1.52_{-0.31}^{+0.49}$ & -3.94$\pm$0.15 & 128.39$\pm$0.96$^{2}$ \\ 
J04183444+2830302 & SS94 V410 X-ray 2 & -9  & N & ... & 15.40$\pm$2.30 & 8 & M0 & 12 & $-0.25_{-0.10}^{+0.04}$ & $-0.11_{-0.14}^{+0.06}$ & -2.89$\pm$0.10 & 128.85$\pm$1.14$^{2}$ \\ 
J04184133+2827250 & LR1 & -9  & N & ... & 30.80$\pm$1.60 & 8 & K4.5 & 19 & $-0.07_{-0.04}^{+0.08}$ & $0.18_{-0.07}^{+0.13}$ & -2.67$\pm$0.09 & 128.85$\pm$1.14$^{2}$ \\ 
J04184250+2818498 & SS94 V410 X-ray 7 AB & -1  & N & ... & $<$13.00 & 8 & M0.5/2.75 & 15/8 & $0.01_{-0.10}^{+0.07}$ & $0.01_{-0.16}^{+0.10}$ & $<$-3.02 & 125.27$\pm$4.07 \\ 
J04190110+2819420 & V410 X-ray 6 & 0  & Y & ... & $<$0.16 & 11 & M5.9 & 14 & $-1.24_{-0.16}^{+0.33}$ & $-1.56_{-0.25}^{+0.57}$ & $<$-4.59 & 119.03$\pm$2.19 \\ 
J04190126+2802487 & KPNO 12  & 0  & N & ... & $<$2.10 & 8 & M9.25 & 22 & $-1.77_{-0.13}^{+0.14}$ & $-2.35_{-0.15}^{+0.21}$ & $<$-3.20 & 128.36$\pm$1.09$^{2}$ \\ 
J04191281+2829330 & FQ Tau A & 1  & Y &  0.780$^{1}$ & 3.10$\pm$0.21 & 9 & M3 & 16 & $-0.54_{-0.19}^{+0.10}$ & $-0.50_{-0.19}^{+0.12}$ & -3.49$\pm$0.09 & 128.85$\pm$1.14$^{2}$ \\ 
J04191281+2829330 & FQ Tau B & 2  & Y &  0.780$^{1}$ & 2.70$\pm$0.15 & 9 & M3.5 & 16 & $-0.62_{-0.20}^{+0.14}$ & $-0.59_{-0.26}^{+0.16}$ & -3.53$\pm$0.09 & 128.85$\pm$1.14$^{2}$ \\ 
J04191583+2906269 & BP Tau  & 0  & Y & ... & 41.50$\pm$2.20 & 8 & M0.5 & 14 & $-0.29_{-0.10}^{+0.07}$ & $-0.17_{-0.15}^{+0.10}$ & -2.45$\pm$0.09 & 128.61$\pm$0.96 \\ 
J04192625+2826142 & V819 Tau & 0  & Y & ... & 0.53$\pm$0.14 & 10 & K8 & 14 & $-0.20_{-0.01}^{+0.01}$ & $-0.03_{-0.01}^{+0.01}$ & -4.36$\pm$0.12 & 131.22$\pm$1.10 \\ 
J04193545+2827218 & FR Tau & -9  & N & ... & $<$15.00 & 8 & M5.3 & 14 & $-1.00_{-0.27}^{+0.21}$ & $-1.17_{-0.42}^{+0.41}$ & $<$-2.64 & 128.39$\pm$0.96$^{2}$ \\ 
J04201611+2821325 &  J04201611+2821325 & -9  & N & ... & $<$1.70 & 8 & M6.5 & 12 & $-1.33_{-0.17}^{+0.27}$ & $-1.70_{-0.25}^{+0.45}$ & $<$-3.46 & 128.66$\pm$3.35 \\ 
J04202144+2813491 & J04202144+2813491 & -9  & N & ... & 52.40$\pm$1.50 & 8 & M1 & 12 & $-0.35_{-0.09}^{+0.10}$ & $-0.25_{-0.13}^{+0.14}$ & -2.35$\pm$0.09 & 126.23$\pm$1.59$^{2}$ \\ 
J04202555+2700355 & J04202555+2700355 & 0  & Y & ... & 5.79$\pm$0.16 & 11 & M5.25 & 13 & $-0.98_{-0.27}^{+0.21}$ & $-1.14_{-0.43}^{+0.40}$ & -2.82$\pm$0.13 & 169.80$\pm$5.38 \\ 
J04202583+2819237 & IRAS 04173+2812 & -9  & N & ... & $<$2.00 & 8 & M4 & 12 & $-0.73_{-0.19}^{+0.19}$ & $-0.68_{-0.35}^{+0.19}$ & $<$-3.68 & 122.27$\pm$6.43 \\ 
J04202606+2804089 &  J04202606+2804089 & -9  & N & ... & $<$4.30 & 8 & M3.5 & 14 & $-0.62_{-0.20}^{+0.14}$ & $-0.59_{-0.26}^{+0.16}$ & $<$-3.34 & 127.05$\pm$0.86 \\ 
J04210795+2702204 & CFHT-BD-Tau 19  & 0  & Y & ... & $<$2.70 & 8 & M5.25 & 12 & $-0.98_{-0.27}^{+0.21}$ & $-1.14_{-0.43}^{+0.40}$ & $<$-3.20 & 160.31$\pm$2.41$^{2}$ \\ 
J04210934+2750368 &  J04210934+2750368 & -9  & N & ... & $<$1.80 & 8 & M4 & 23 & $-0.73_{-0.19}^{+0.19}$ & $-0.68_{-0.35}^{+0.19}$ & $<$-3.76 & 117.94$\pm$1.37 \\ 
J04213459+2701388 & J04213459+2701388 & 0  & Y & ... & $<$0.17 & 11 & M5.5 & 13 & $-1.05_{-0.27}^{+0.23}$ & $-1.25_{-0.45}^{+0.39}$ & $<$-4.35 & 166.34$\pm$3.85 \\ 
J04214323+1934133 & IRAS 04187+1927 & 0  & Y & ... & 3.70$\pm$0.21 & 7 & M2.4 & 14 & $-0.47_{-0.13}^{+0.08}$ & $-0.42_{-0.15}^{+0.11}$ & -3.31$\pm$0.09 & 148.16$\pm$2.19 \\ 
J04214631+2659296 &  J04214631+2659296 & 0  & Y & ... & $<$3.90 & 8 & M5.75 & 17 & $-1.25_{-0.13}^{+0.38}$ & $-1.58_{-0.20}^{+0.65}$ & $<$-2.93 & 160.29$\pm$7.35 \\ 
J04215563+2755060 & DE Tau & 0  & Y & ... & 31.10$\pm$3.10 & 8 & M2.3 & 14 & $-0.46_{-0.12}^{+0.09}$ & $-0.41_{-0.14}^{+0.12}$ & -2.53$\pm$0.09 & 126.92$\pm$1.07 \\ 
J04215740+2826355 & RY Tau  & 0  & Y & ... & 192.50$\pm$9.10 & 8 & G0 & 14 & $0.54_{-0.01}^{+0.01}$ & $1.72_{-0.08}^{+0.05}$ & -2.23$\pm$0.08 & 132.82$\pm$2.57$^{2}$ \\ 
J04220217+2657304 & FS Tau A & 1  & Y &  0.227$^{1}$ & 2.27$\pm$0.18 & 7 & M0 & 16 & $-0.25_{-0.10}^{+0.04}$ & $-0.11_{-0.14}^{+0.06}$ & -3.54$\pm$0.09 & 160.31$\pm$2.41$^{2}$ \\ 
J04220217+2657304 & FS Tau B & 2  & Y &  0.227$^{1}$ & $<$0.52 & 7 & M3.5 & 16 & $-0.62_{-0.20}^{+0.14}$ & $-0.59_{-0.26}^{+0.16}$ & $<$-4.06 & 160.31$\pm$2.41$^{2}$ \\ 
J04221675+2654570 & CFHT-Tau-21 & 0  & Y & ... & $<$4.20 & 8 & M1.5 & 14 & $-0.40_{-0.08}^{+0.10}$ & $-0.32_{-0.11}^{+0.15}$ & $<$-3.23 & 156.97$\pm$3.36 \\ 
J04224786+2645530 & IRAS 04196+2638 & -9  & N & ... & 51.00$\pm$1.20 & 8 & M1 & 24 & $-0.35_{-0.09}^{+0.10}$ & $-0.25_{-0.13}^{+0.14}$ & -2.17$\pm$0.09 & 157.03$\pm$2.73 \\ 
J04230607+2801194 & J04230607+2801194 & 0  & Y & ... & 2.28$\pm$0.14 & 11 & M6 & 12 & $-1.22_{-0.20}^{+0.29}$ & $-1.52_{-0.31}^{+0.49}$ & -3.34$\pm$0.14 & 133.44$\pm$2.45 \\ 
J04230776+2805573 & IRAS 04200+2759  & -9  & N & ... & 36.60$\pm$1.30 & 8 & M2 & 12 & $-0.44_{-0.10}^{+0.09}$ & $-0.38_{-0.12}^{+0.13}$ & -2.39$\pm$0.09 & 138.64$\pm$3.34 \\ 
J04231822+2641156 &  J04231822+2641156 & -9  & N & ... & $<$3.90 & 8 & M3.5 & 24 & $-0.62_{-0.20}^{+0.14}$ & $-0.59_{-0.26}^{+0.16}$ & $<$-3.22 & 152.80$\pm$3.88$^{2}$ \\ 
J04233539+2503026 & FU Tau A & 1  & N & ... & $<$1.70 & 8 & M7.25 & 25 & $-1.45_{-0.18}^{+0.20}$ & $-1.89_{-0.26}^{+0.32}$ & $<$-3.39 & 131.20$\pm$2.60 \\ 
J04233539+2503026 & FU Tau B & 2  & N & ... & $<$1.70 & 8 & M9.25 & 25 & $-1.77_{-0.13}^{+0.14}$ & $-2.35_{-0.15}^{+0.21}$ & $<$-3.27 & 131.20$\pm$2.60$^{1}$ \\ 
J04233919+2456141 & FT Tau & 0  & Y & ... & 62.90$\pm$4.30 & 8 & M2.8 & 14 & $-0.51_{-0.18}^{+0.09}$ & $-0.47_{-0.19}^{+0.11}$ & -2.20$\pm$0.09 & 127.34$\pm$0.85 \\ 
J04242090+2630511 &  J04242090+2630511 & -9  & N & ... & $<$1.40 & 8 & M6.5 & 24 & $-1.33_{-0.17}^{+0.27}$ & $-1.70_{-0.25}^{+0.45}$ & $<$-3.48 & 137.19$\pm$3.48 \\ 
J04242646+2649503 &  J04242646+2649503 & 0  & Y & ... & $<$3.00 & 8 & M5.75 & 17 & $-1.25_{-0.13}^{+0.38}$ & $-1.58_{-0.20}^{+0.65}$ & $<$-3.08 & 154.92$\pm$3.15 \\ 
J04244457+2610141 & IRAS 04216+2603 & -9  & N & ... & 19.80$\pm$2.00 & 8 & M2.8 & 14 & $-0.51_{-0.18}^{+0.09}$ & $-0.47_{-0.19}^{+0.11}$ & -2.51$\pm$0.09 & 158.92$\pm$2.77 \\ 
J04245708+2711565 & IP Tau  & 0  & Y & ... & 8.80$\pm$1.50 & 8 & M0.6 & 14 & $-0.30_{-0.10}^{+0.08}$ & $-0.18_{-0.15}^{+0.11}$ & -3.11$\pm$0.10 & 130.09$\pm$0.72 \\ 
J04262939+2624137 & KPNO 3 & 0  & Y & ... & 2.29$\pm$0.09 & 11 & M6 & 12 & $-1.22_{-0.20}^{+0.29}$ & $-1.52_{-0.31}^{+0.49}$ & -3.21$\pm$0.14 & 155.51$\pm$5.55 \\ 
J04263055+2443558 &  J04263055+2443558 & 0  & N & ... & $<$0.30 & 7 & M8.75 & 17 & $-1.70_{-0.13}^{+0.15}$ & $-2.26_{-0.18}^{+0.24}$ & $<$-4.13 & 119.88$\pm$10.05 \\ 
J04265352+2606543 & FV Tau A & 1  & Y &  0.720$^{1}$ & 6.17$\pm$0.16 & 9 & K5 & 16 & $-0.10_{-0.05}^{+0.06}$ & $0.13_{-0.08}^{+0.12}$ & -3.26$\pm$0.09 & 143.02$\pm$6.31$^{2}$ \\ 
J04265352+2606543 & FV Tau B & 2  & Y &  0.720$^{1}$ & 5.93$\pm$0.18 & 9 & K6 & 1 & $-0.15_{-0.04}^{+0.05}$ & $0.05_{-0.07}^{+0.08}$ & -3.26$\pm$0.09 & 143.02$\pm$6.31$^{2}$ \\ 
J04265440+2606510 & FV Tau/c A & 1  & Y &  0.740$^{1}$ & 0.76$\pm$0.17 & 9 & M2.5 & 12 & $-0.48_{-0.14}^{+0.08}$ & $-0.43_{-0.16}^{+0.11}$ & -4.05$\pm$0.12 & 139.43$\pm$2.65 \\ 
J04265440+2606510 & FV Tau/c B & 2  & Y &  0.740$^{1}$ & 0.40$\pm$0.12 & 9 & M3.5 & 16 & $-0.62_{-0.20}^{+0.14}$ & $-0.59_{-0.26}^{+0.16}$ & -4.29$\pm$0.14 & 139.43$\pm$2.65$^{1}$ \\ 
J04265732+2606284 & KPNO 13  & 0  & Y & ... & $<$3.60 & 8 & M5.1 & 14 & $-0.94_{-0.28}^{+0.20}$ & $-1.07_{-0.45}^{+0.37}$ & $<$-3.26 & 132.92$\pm$2.27 \\ 
J04270280+2542223 & DF Tau AB & -1  & N & ... & 3.40$\pm$1.70 & 8 & M2/2.5 & 16 & $-0.14_{-0.10}^{+0.09}$ & $-0.10_{-0.14}^{+0.12}$ & -3.51$\pm$0.19 & 135.66$\pm$3.22$^{2}$ \\ 
J04270469+2606163 & DG Tau & 0  & Y & ... & 344.20$\pm$17.80 & 8 & K7.0 & 14 & $-0.19_{-0.01}^{+0.04}$ & $-0.02_{-0.01}^{+0.07}$ & -1.51$\pm$0.09 & 137.41$\pm$4.53$^{2}$ \\ 
J04284263+2714039 & J04284263+2714039 A & 1  & Y &  0.630$^{2}$ & 0.67$\pm$0.10 & 11 & M5.25 & 13 & $-0.98_{-0.27}^{+0.21}$ & $-1.14_{-0.43}^{+0.40}$ & -3.97$\pm$0.14 & 132.37$\pm$2.53 \\ 
J04284263+2714039 & J04284263+2714039 B & 2  & Y &  0.630$^{2}$ & $<$0.29 & 11 & M5.25 & 13 & $-0.98_{-0.27}^{+0.21}$ & $-1.14_{-0.43}^{+0.40}$ & $<$-4.34 & 132.37$\pm$2.53$^{1}$ \\ 
J04290068+2755033 &  J04290068+2755033 & 0  & N & ... & $<$1.60 & 8 & M8.25 & 17 & $-1.63_{-0.14}^{+0.18}$ & $-2.14_{-0.21}^{+0.26}$ & $<$-3.26 & 145.71$\pm$7.67 \\ 
J04290498+2649073 & IRAS 04260+2642 & -9  & N & ... & 120.00$\pm$10.00 & 8 & K5 & 26 & $-0.10_{-0.05}^{+0.06}$ & $0.13_{-0.08}^{+0.12}$ & -2.03$\pm$0.09 & 134.35$\pm$0.99$^{2}$ \\ 
J04292165+2701259 & IRAS 04263+2654  & 0  & Y & ... & 2.89$\pm$0.14 & 11 & M5.25 & 17 & $-0.98_{-0.27}^{+0.21}$ & $-1.14_{-0.43}^{+0.40}$ & -3.36$\pm$0.13 & 129.11$\pm$2.37$^{2}$ \\ 
J04293606+2435556 & XEST 13-010 & -9  & N & ... & 15.20$\pm$1.20 & 8 & M3 & 12 & $-0.54_{-0.19}^{+0.10}$ & $-0.50_{-0.19}^{+0.12}$ & -2.80$\pm$0.09 & 128.97$\pm$1.50 \\ 
J04294155+2632582 & DH Tau A & 1  & N & ... & 18.00$\pm$8.20 & 8 & M1 & 20 & $-0.35_{-0.09}^{+0.10}$ & $-0.25_{-0.13}^{+0.14}$ & -2.75$\pm$0.18 & 134.85$\pm$1.27 \\ 
J04294155+2632582 & DH Tau B & 2  & N & ... & $<$3.80 & 8 & M7.5 & 8 & $-1.50_{-0.17}^{+0.17}$ & $-1.95_{-0.25}^{+0.25}$ & $<$-3.00 & 134.85$\pm$1.27$^{1}$ \\ 
J04295156+2606448 & IQ Tau  & 0  & Y & ... & 61.90$\pm$4.50 & 8 & M1.1 & 14 & $-0.36_{-0.09}^{+0.10}$ & $-0.26_{-0.13}^{+0.14}$ & -2.24$\pm$0.09 & 130.79$\pm$1.06 \\ 
J04295950+2433078 & CFHT-BD-Tau 20 & 0  & Y & ... & 2.90$\pm$0.17 & 7 & M5 & 17 & $-0.92_{-0.29}^{+0.19}$ & $-1.03_{-0.49}^{+0.35}$ & -3.38$\pm$0.12 & 130.72$\pm$2.63 \\ 
J04300399+1813493 & UX Tau A & 1  & N & ... & 79.00$\pm$3.95 & 7 & G8 & 27 & $0.37_{-0.09}^{+0.05}$ & $0.95_{-0.17}^{+0.13}$ & -2.38$\pm$0.09 & 139.40$\pm$1.96 \\ 
J04300399+1813493 & UX Tau Ba & 2  & N & ... & $<$0.36 & 7 & M2 & 28 & $-0.44_{-0.10}^{+0.09}$ & $-0.38_{-0.12}^{+0.13}$ & $<$-4.39 & 139.40$\pm$1.96$^{1}$ \\ 
J04300399+1813493 & UX Tau Bb & 3  & N & ... & $<$0.36 & 7 & M3.0 & 7 & $-0.54_{-0.19}^{+0.10}$ & $-0.50_{-0.19}^{+0.12}$ & $<$-4.36 & 139.40$\pm$1.96$^{1}$ \\ 
J04300399+1813493 & UX Tau C & 4  & N & ... & $<$0.36 & 7 & M2.8 & 14 & $-0.51_{-0.18}^{+0.09}$ & $-0.47_{-0.19}^{+0.11}$ & $<$-4.37 & 139.40$\pm$1.96$^{1}$ \\ 
J04300724+2608207 & KPNO 6  & 0  & N & ... & $<$2.40 & 8 & M8.5 & 19 & $-1.67_{-0.13}^{+0.17}$ & $-2.20_{-0.19}^{+0.25}$ & $<$-3.27 & 116.47$\pm$7.46 \\ 
J04302961+2426450 & FX Tau A & 1  & Y &  0.890$^{1}$ & 7.84$\pm$0.53 & 7 & M1 & 28 & $-0.35_{-0.09}^{+0.10}$ & $-0.25_{-0.13}^{+0.14}$ & -3.14$\pm$0.09 & 130.11$\pm$0.64$^{2}$ \\ 
J04302961+2426450 & FX Tau B & 2  & Y &  0.890$^{1}$ & $<$0.36 & 7 & M4 & 28 & $-0.73_{-0.19}^{+0.19}$ & $-0.68_{-0.35}^{+0.19}$ & $<$-4.37 & 130.11$\pm$0.64$^{2}$ \\ 
J04304425+2601244 & DK Tau A & 1  & Y &  2.360$^{3}$ & 30.30$\pm$0.18 & 9 & K8.5 & 14 & $-0.20_{-0.02}^{+0.00}$ & $-0.04_{-0.03}^{+0.01}$ & -2.62$\pm$0.08 & 128.05$\pm$0.98 \\ 
J04304425+2601244 & DK Tau B & 2  & Y &  2.360$^{3}$ & 2.88$\pm$0.19 & 9 & M1.7 & 14 & $-0.42_{-0.09}^{+0.10}$ & $-0.34_{-0.11}^{+0.14}$ & -3.57$\pm$0.09 & 128.05$\pm$0.98$^{1}$ \\ 
J04305137+2442222 & ZZ Tau AB & -1  & N & ... & 0.59$\pm$0.11 & 7 & M3/4.5 & 23/8 & $-0.24_{-0.19}^{+0.10}$ & $-0.34_{-0.24}^{+0.17}$ & -4.24$\pm$0.11 & 130.73$\pm$1.26$^{2}$ \\ 
J04305171+2441475 & ZZ Tau IRS & 0  & Y & ... & 105.80$\pm$1.50 & 8 & M4.5 & 14 & $-0.83_{-0.23}^{+0.20}$ & $-0.85_{-0.39}^{+0.26}$ & -1.86$\pm$0.10 & 130.73$\pm$1.26$^{2}$ \\ 
J04305718+2556394 & KPNO 7  & 0  & N & ... & $<$2.60 & 8 & M8.25 & 19 & $-1.63_{-0.14}^{+0.18}$ & $-2.14_{-0.21}^{+0.26}$ & $<$-3.20 & 123.09$\pm$7.18 \\ 
J04311444+2710179 & JH 56 & 0  & Y & ... & $<$3.10 & 8 & K8.0 & 14 & $-0.20_{-0.01}^{+0.01}$ & $-0.03_{-0.01}^{+0.01}$ & $<$-3.62 & 127.02$\pm$0.72 \\ 
J04314007+1813571 & XZ Tau A & 1  & Y &  0.300$^{1}$ & 7.30$\pm$0.42 & 7 & M2 & 16 & $-0.44_{-0.10}^{+0.09}$ & $-0.38_{-0.12}^{+0.13}$ & -3.05$\pm$0.08 & 144.15$\pm$1.55$^{2}$ \\ 
J04314007+1813571 & XZ Tau B & 2  & Y &  0.300$^{1}$ & 8.90$\pm$0.49 & 7 & M3.5 & 16 & $-0.62_{-0.20}^{+0.14}$ & $-0.59_{-0.26}^{+0.16}$ & -2.91$\pm$0.09 & 144.15$\pm$1.55$^{2}$ \\ 
J04315056+2424180 & HK Tau A & 1  & Y &  2.350$^{1}$ & 33.80$\pm$0.20 & 9 & M1 & 29 & $-0.35_{-0.09}^{+0.10}$ & $-0.25_{-0.13}^{+0.14}$ & -2.49$\pm$0.09 & 132.85$\pm$1.64 \\ 
J04315056+2424180 & HK Tau B & 2  & Y &  2.350$^{1}$ & 16.00$\pm$0.24 & 9 & M2 & 29 & $-0.44_{-0.10}^{+0.09}$ & $-0.38_{-0.12}^{+0.13}$ & -2.78$\pm$0.08 & 132.85$\pm$1.64$^{1}$ \\ 
J04315779+1821380 & V710 Tau A & 1  & Y &  3.170$^{1}$ & 66.00$\pm$3.31 & 9 & M1.7 & 14 & $-0.42_{-0.09}^{+0.10}$ & $-0.34_{-0.11}^{+0.14}$ & -2.12$\pm$0.09 & 142.42$\pm$2.28 \\ 
J04315779+1821380 & V710 Tau B & 2  & Y &  3.170$^{1}$ & $<$0.61 & 9 & M3.3 & 14 & $-0.59_{-0.20}^{+0.12}$ & $-0.55_{-0.22}^{+0.14}$ & $<$-4.10 & 142.42$\pm$2.28$^{1}$ \\ 
J04315968+1821305 & LkHa 267 & -9  & N & ... & $<$0.30 & 7 & M1.7 & 14 & $-0.42_{-0.09}^{+0.10}$ & $-0.34_{-0.11}^{+0.14}$ & $<$-4.39 & 154.81$\pm$3.79 \\ 
J04321540+2428597 & Haro 6-13 & 0  & Y & ... & 119.60$\pm$5.70 & 8 & M0 & 30 & $-0.25_{-0.10}^{+0.04}$ & $-0.11_{-0.14}^{+0.06}$ & -2.00$\pm$0.08 & 129.68$\pm$0.51$^{2}$ \\ 
J04321606+1812464 & MHO 5  & 0  & N & ... & $<$0.31 & 7 & M6.5 & 14 & $-1.33_{-0.17}^{+0.27}$ & $-1.70_{-0.25}^{+0.45}$ & $<$-4.09 & 144.06$\pm$2.00 \\ 
J04322210+1827426 & MHO 6 & 0  & Y & ... & 19.37$\pm$0.30 & 11 & M5.0 & 14 & $-0.92_{-0.29}^{+0.19}$ & $-1.03_{-0.49}^{+0.35}$ & -2.48$\pm$0.12 & 141.38$\pm$1.96 \\ 
J04322415+2251083 &  J04322415+2251083 & -9  & N & ... & $<$0.30 & 7 & M4.5 & 12 & $-0.83_{-0.23}^{+0.20}$ & $-0.85_{-0.39}^{+0.26}$ & $<$-4.26 & 154.76$\pm$2.91 \\ 
J04323028+1731303 & GG Tau Aa & 1  & N & ... & 7.00$\pm$0.87 & 7 & K7 & 31 & $-0.19_{-0.01}^{+0.04}$ & $-0.02_{-0.01}^{+0.07}$ & -3.13$\pm$0.09 & 149.46$\pm$2.21 \\ 
J04323028+1731303 & GG Tau Ab & 2  & N & ... & $<$2.40 & 7 & M0.5 & 31 & $-0.29_{-0.10}^{+0.07}$ & $-0.17_{-0.15}^{+0.10}$ & $<$-3.56 & 149.46$\pm$2.21$^{1}$ \\ 
J04323028+1731303 & GG Tau Ba & 3  & N & ... & $<$2.40 & 7 & M5.5 & 31 & $-1.05_{-0.27}^{+0.23}$ & $-1.25_{-0.45}^{+0.39}$ & $<$-3.29 & 149.46$\pm$2.21$^{1}$ \\ 
J04323028+1731303 & GG Tau Bb & 4  & N & ... & $<$2.40 & 7 & M7.5 & 31 & $-1.50_{-0.17}^{+0.17}$ & $-1.95_{-0.25}^{+0.25}$ & $<$-3.11 & 149.46$\pm$2.21$^{1}$ \\ 
J04323058+2419572 & FY Tau  & 1  & Y & 17.200$^{4}$ & 13.80$\pm$5.10 & 8 & M0.1 & 14 & $-0.26_{-0.10}^{+0.05}$ & $-0.12_{-0.14}^{+0.07}$ & -2.93$\pm$0.15 & 129.74$\pm$1.20 \\ 
J04323176+2420029 & FZ Tau  & 2  & Y & 17.200$^{4}$ & 13.70$\pm$2.40 & 8 & M0.5 & 14 & $-0.29_{-0.10}^{+0.07}$ & $-0.17_{-0.15}^{+0.10}$ & -2.92$\pm$0.11 & 129.56$\pm$1.26 \\ 
J04324303+2552311 & UZ Tau Eab & -1  & N & ... & 128.10$\pm$7.30 & 8 & M1/4 & 20/32 & $-0.05_{-0.09}^{+0.10}$ & $-0.11_{-0.18}^{+0.15}$ & -1.94$\pm$0.10 & 134.49$\pm$7.67$^{2}$ \\ 
J04324303+2552311 & UZ Tau Wa & 2  & N & ... & 16.20$\pm$3.90 & 8 & M2 & 16 & $-0.44_{-0.10}^{+0.09}$ & $-0.38_{-0.12}^{+0.13}$ & -2.77$\pm$0.13 & 134.49$\pm$7.67$^{2}$ \\ 
J04324303+2552311 & UZ Tau Wb & 3  & N & ... & 31.00$\pm$4.10 & 8 & M3 & 16 & $-0.54_{-0.19}^{+0.10}$ & $-0.50_{-0.19}^{+0.12}$ & -2.46$\pm$0.11 & 134.49$\pm$7.67$^{2}$ \\ 
J04324911+2253027 & IRAS04298+2246Aa & 1  & Y &  1.650$^{5}$ & 3.53$\pm$0.23 & 9 & K5.5 & 14 & $-0.11_{-0.07}^{+0.04}$ & $0.11_{-0.10}^{+0.07}$ & -3.38$\pm$0.09 & 163.79$\pm$2.14 \\ 
J04324911+2253027 & IRAS04298+2246Ab & 2  & Y &  1.650$^{5}$ & 3.06$\pm$0.24 & 9 & M4.6 & 14 & $-0.84_{-0.22}^{+0.20}$ & $-0.89_{-0.36}^{+0.28}$ & -3.19$\pm$0.11 & 163.79$\pm$2.14$^{1}$ \\ 
J04324911+2253027 & IRAS04298+2246Ba & 1  & N & ... & 0.25$\pm$0.08 & 9 & M8.5 & 7 & $-1.67_{-0.13}^{+0.17}$ & $-2.20_{-0.19}^{+0.25}$ & -3.95$\pm$0.15 & 163.79$\pm$2.14$^{1}$ \\ 
J04324911+2253027 & IRAS04298+2246Bb & 2  & N & ... & 2.30$\pm$0.24 & 9 & M8.5 & 7 & $-1.67_{-0.13}^{+0.17}$ & $-2.20_{-0.19}^{+0.25}$ & -2.99$\pm$0.10 & 163.79$\pm$2.14$^{1}$ \\ 
J04330622+2409339 & GH Tau  A & 1  & Y &  0.310$^{1}$ & 3.90$\pm$0.22 & 7 & M2 & 16 & $-0.44_{-0.10}^{+0.09}$ & $-0.38_{-0.12}^{+0.13}$ & -3.42$\pm$0.08 & 129.68$\pm$0.51$^{2}$ \\ 
J04330622+2409339 & GH Tau B & 2  & Y &  0.310$^{1}$ & 2.90$\pm$0.18 & 7 & M2 & 16 & $-0.44_{-0.10}^{+0.09}$ & $-0.38_{-0.12}^{+0.13}$ & -3.55$\pm$0.09 & 129.68$\pm$0.51$^{2}$ \\ 
J04330664+2409549 & V807 Tau A & 1  & N & ... & 8.90$\pm$0.46 & 7 & K7 & 33 & $-0.19_{-0.01}^{+0.04}$ & $-0.02_{-0.01}^{+0.07}$ & -3.15$\pm$0.08 & 129.68$\pm$0.51$^{2}$ \\ 
J04330664+2409549 & V807 Tau Bab & -2  & N & ... & $<$0.33 & 7 & M2/2.5 & 33 & $-0.14_{-0.10}^{+0.09}$ & $-0.10_{-0.14}^{+0.12}$ & $<$-4.56 & 129.68$\pm$0.51$^{2}$ \\ 
J04330945+2246487 &  J04330945+2246487 & 0  & Y & ... & $<$0.28 & 7 & M6 & 17 & $-1.22_{-0.20}^{+0.29}$ & $-1.52_{-0.31}^{+0.49}$ & $<$-4.08 & 160.63$\pm$3.81$^{2}$ \\ 
J04331435+2614235 & IRAS 04301+2608 & -9  & N & ... & 6.60$\pm$3.70 & 8 & M0 & 12 & $-0.25_{-0.10}^{+0.04}$ & $-0.11_{-0.14}^{+0.06}$ & -2.72$\pm$0.26 & 240.02$\pm$65.31 \\ 
J04331907+2246342 & IRAS 04303+2240 & -9  & N & ... & $<$6.00 & 8 & M0.5 & 30 & $-0.29_{-0.10}^{+0.07}$ & $-0.17_{-0.15}^{+0.10}$ & $<$-3.10 & 160.63$\pm$3.81$^{2}$ \\ 
J04333278+1800436 & J04333278+1800436 & -9  & N & ... & 11.00$\pm$1.90 & 8 & M1 & 34 & $-0.35_{-0.09}^{+0.10}$ & $-0.25_{-0.13}^{+0.14}$ & -2.89$\pm$0.11 & 146.59$\pm$1.25 \\ 
J04333405+2421170 & GI Tau & 2  & Y & 13.100$^{6}$ & 12.00$\pm$1.00 & 8 & M0.4 & 14 & $-0.28_{-0.10}^{+0.07}$ & $-0.16_{-0.15}^{+0.09}$ & -2.98$\pm$0.09 & 130.02$\pm$0.78 \\ 
J04333456+2421058 & GK Tau & 1  & Y & 13.100$^{6}$ & 5.20$\pm$0.22 & 9 & K6.5 & 14 & $-0.18_{-0.02}^{+0.07}$ & $0.00_{-0.03}^{+0.10}$ & -3.39$\pm$0.08 & 128.79$\pm$0.73 \\ 
J04333678+2609492 & IS Tau A & 1  & Y &  0.220$^{1}$ & 1.15$\pm$0.16 & 7 & M0 & 16 & $-0.25_{-0.10}^{+0.04}$ & $-0.11_{-0.14}^{+0.06}$ & -3.83$\pm$0.10 & 161.35$\pm$2.37$^{2}$ \\ 
J04333678+2609492 & IS Tau B & 2  & Y &  0.220$^{1}$ & 1.05$\pm$0.16 & 7 & M3.5 & 16 & $-0.62_{-0.20}^{+0.14}$ & $-0.59_{-0.26}^{+0.16}$ & -3.74$\pm$0.10 & 161.35$\pm$2.37$^{2}$ \\ 
J04333905+2227207 & J04333905+2227207 & -9  & N & ... & 31.00$\pm$2.00 & 8 & M1.75 & 12 & $-0.42_{-0.09}^{+0.10}$ & $-0.35_{-0.11}^{+0.14}$ & -0.93$\pm$0.27 & 810.16$\pm$346.28 \\ 
J04333906+2520382 & DL Tau  & 0  & Y & ... & 168.80$\pm$10.80 & 8 & K5.5 & 14 & $-0.11_{-0.07}^{+0.04}$ & $0.11_{-0.10}^{+0.07}$ & -1.73$\pm$0.08 & 158.62$\pm$1.21 \\ 
J04333935+1751523 & HN Tau A & 1  & Y &  3.140$^{3}$ & 15.70$\pm$0.79 & 7 & K3 & 14 & $0.05_{-0.08}^{+0.07}$ & $0.38_{-0.14}^{+0.11}$ & -2.90$\pm$0.08 & 145.68$\pm$0.74$^{2}$ \\ 
J04333935+1751523 & HN Tau B & 2  & Y &  3.140$^{3}$ & 0.54$\pm$0.10 & 7 & M4.8 & 14 & $-0.88_{-0.37}^{+0.19}$ & $-0.95_{-0.64}^{+0.30}$ & -4.03$\pm$0.13 & 145.68$\pm$0.74$^{2}$ \\ 
J04334171+1750402 & J04334171+1750402 & -9  & N & ... & 6.00$\pm$2.00 & 8 & M4 & 35 & $-0.73_{-0.19}^{+0.19}$ & $-0.68_{-0.35}^{+0.19}$ & -3.05$\pm$0.15 & 145.74$\pm$1.54 \\ 
J04334465+2615005 & J04334465+2615005 & -9  & N & ... & 14.11$\pm$0.27 & 11 & M5.2 & 14 & $-0.97_{-0.27}^{+0.20}$ & $-1.12_{-0.43}^{+0.39}$ & -2.48$\pm$0.12 & 161.35$\pm$2.37$^{2}$ \\ 
J04334871+1810099 & DM Tau  & 0  & Y & ... & 89.40$\pm$3.10 & 8 & M3.0 & 14 & $-0.54_{-0.19}^{+0.10}$ & $-0.50_{-0.19}^{+0.12}$ & -1.94$\pm$0.09 & 144.55$\pm$1.08 \\ 
J04335470+2613275 & IT Tau A & 1  & Y &  2.410$^{3}$ & 7.00$\pm$0.24 & 9 & K6.0 & 14 & $-0.15_{-0.04}^{+0.05}$ & $0.05_{-0.07}^{+0.08}$ & -3.08$\pm$0.08 & 161.27$\pm$1.97 \\ 
J04335470+2613275 & IT Tau B & 2  & Y &  2.410$^{3}$ & 4.17$\pm$0.27 & 9 & M2.9 & 14 & $-0.52_{-0.19}^{+0.09}$ & $-0.48_{-0.19}^{+0.11}$ & -3.17$\pm$0.09 & 161.27$\pm$1.97$^{1}$ \\ 
J04345542+2428531 & AA Tau  & 0  & Y & ... & 65.00$\pm$3.50 & 8 & M0.6 & 14 & $-0.30_{-0.10}^{+0.08}$ & $-0.18_{-0.15}^{+0.11}$ & -2.25$\pm$0.09 & 128.78$\pm$0.67$^{2}$ \\ 
J04352020+2232146 & HO Tau & 0  & Y & ... & 16.20$\pm$0.20 & 9 & M3.2 & 14 & $-0.57_{-0.20}^{+0.11}$ & $-0.53_{-0.20}^{+0.13}$ & -2.58$\pm$0.09 & 160.67$\pm$1.24 \\ 
J04352737+2414589 & DN Tau  & 0  & Y & ... & 82.30$\pm$4.50 & 8 & M0.3 & 14 & $-0.27_{-0.11}^{+0.06}$ & $-0.14_{-0.15}^{+0.09}$ & -2.16$\pm$0.09 & 127.76$\pm$0.89 \\ 
J04354093+2411087 & CoKu Tau 3 A & 1  & Y &  2.100$^{1}$ & 1.84$\pm$0.30 & 9 & M0.5 & 14 & $-0.29_{-0.10}^{+0.07}$ & $-0.17_{-0.15}^{+0.10}$ & -3.83$\pm$0.10 & 124.85$\pm$2.28 \\ 
J04354093+2411087 & CoKu Tau 3 B & 2  & Y &  2.100$^{1}$ & 5.79$\pm$0.29 & 9 & M4.3 & 14 & $-0.79_{-0.21}^{+0.20}$ & $-0.77_{-0.41}^{+0.22}$ & -3.18$\pm$0.10 & 124.85$\pm$2.28$^{1}$ \\ 
J04354733+2250216 & HQ Tau & 0  & Y & ... & 4.20$\pm$2.20 & 8 & K2.0 & 14 & $0.11_{-0.07}^{+0.04}$ & $0.50_{-0.11}^{+0.07}$ & -3.41$\pm$0.19 & 162.54$\pm$5.08$^{2}$ \\ 
J04355277+2254231 & HP Tau & 0  & Y & ... & 51.70$\pm$4.70 & 8 & K4.0 & 14 & $-0.04_{-0.06}^{+0.08}$ & $0.24_{-0.12}^{+0.14}$ & -2.18$\pm$0.09 & 176.35$\pm$3.37 \\ 
J04355684+2254360 & Haro 6-28 A & 1  & Y &  0.650$^{1}$ & 5.10$\pm$0.29 & 7 & M3.1 & 14 & $-0.55_{-0.20}^{+0.10}$ & $-0.51_{-0.19}^{+0.12}$ & -3.07$\pm$0.09 & 162.54$\pm$5.08$^{2}$ \\ 
J04355684+2254360 & Haro 6-28 B & 2  & Y &  0.650$^{1}$ & 0.78$\pm$0.15 & 7 & M3.5 & 16 & $-0.62_{-0.20}^{+0.14}$ & $-0.59_{-0.26}^{+0.16}$ & -3.87$\pm$0.11 & 162.54$\pm$5.08$^{2}$ \\ 
J04361030+2159364 &  J04361030+2159364 & 0  & N & ... & $<$0.31 & 7 & M8.5 & 17 & $-1.67_{-0.13}^{+0.17}$ & $-2.20_{-0.19}^{+0.25}$ & $<$-4.14 & 118.89$\pm$8.68 \\ 
J04362151+2351165 &  J04362151+2351165 & -9  & N & ... & 1.45$\pm$0.12 & 7 & M5.1 & 14 & $-0.94_{-0.28}^{+0.20}$ & $-1.07_{-0.45}^{+0.37}$ & -3.78$\pm$0.12 & 115.12$\pm$1.70 \\ 
J04375670+2546229 & ITG 1 & -9  & N & ... & $<$2.30 & 8 & M6 & 36 & $-1.22_{-0.20}^{+0.29}$ & $-1.52_{-0.31}^{+0.49}$ & $<$-1.71 & 873.64$\pm$95.61 \\ 
J04381486+2611399 & J04381486+2611399 & 0  & N & ... & 0.67$\pm$0.10 & 11 & M7.25 & 13 & $-1.45_{-0.18}^{+0.20}$ & $-1.89_{-0.26}^{+0.32}$ & -3.70$\pm$0.14 & 147.48$\pm$16.01 \\ 
J04382134+2609137 & GM Tau & 0  & Y & ... & 1.05$\pm$0.07 & 11 & M5.0 & 14 & $-0.92_{-0.29}^{+0.19}$ & $-1.03_{-0.49}^{+0.35}$ & -3.77$\pm$0.12 & 137.87$\pm$2.86 \\ 
J04382858+2610494 & DO Tau  & 0  & Y & ... & 108.20$\pm$6.90 & 8 & M0.3 & 14 & $-0.27_{-0.11}^{+0.06}$ & $-0.14_{-0.15}^{+0.09}$ & -1.97$\pm$0.09 & 138.83$\pm$1.02 \\ 
J04385859+2336351 & J04385859+2336351 & 0  & Y & ... & 10.73$\pm$0.18 & 11 & M4.25 & 37 & $-0.78_{-0.21}^{+0.20}$ & $-0.75_{-0.40}^{+0.21}$ & -2.91$\pm$0.10 & 126.40$\pm$1.92 \\ 
J04390163+2336029 & J04390163+2336029 & 0  & Y & ... & 0.45$\pm$0.10 & 11 & M4.9 & 14 & $-0.90_{-0.33}^{+0.19}$ & $-0.99_{-0.57}^{+0.32}$ & -4.22$\pm$0.14 & 127.36$\pm$1.30 \\ 
J04390396+2544264 & J04390396+2544264 & 0  & N & ... & 0.93$\pm$0.12 & 11 & M7.25 & 13 & $-1.45_{-0.18}^{+0.20}$ & $-1.89_{-0.26}^{+0.32}$ & -3.57$\pm$0.12 & 143.62$\pm$4.39 \\ 
J04391741+2247533 & VY Tau A & 1  & Y &  0.660$^{1}$ & 1.90$\pm$0.22 & 7 & M1.5 & 14 & $-0.40_{-0.08}^{+0.10}$ & $-0.32_{-0.11}^{+0.15}$ & -3.57$\pm$0.09 & 158.15$\pm$1.21$^{2}$ \\ 
J04391741+2247533 & VY Tau B & 2  & Y &  0.660$^{1}$ & $<$0.60 & 7 & M4.5 & 8 & $-0.83_{-0.23}^{+0.20}$ & $-0.85_{-0.39}^{+0.26}$ & $<$-3.94 & 158.15$\pm$1.21$^{2}$ \\ 
J04391779+2221034 & LkCa 15  & 0  & Y & ... & 127.00$\pm$4.90 & 8 & K5.5 & 14 & $-0.11_{-0.07}^{+0.04}$ & $0.11_{-0.10}^{+0.07}$ & -1.85$\pm$0.08 & 158.15$\pm$1.21 \\ 
J04392090+2545021 & GN Tau A & 1  & Y &  0.340$^{1}$ & 0.47$\pm$0.12 & 7 & M2.5 & 14 & $-0.48_{-0.14}^{+0.08}$ & $-0.43_{-0.16}^{+0.11}$ & -4.26$\pm$0.13 & 139.23$\pm$1.85$^{2}$ \\ 
J04392090+2545021 & GN Tau B & 2  & Y &  0.340$^{1}$ & 0.59$\pm$0.12 & 7 & M2.5 & 8 & $-0.48_{-0.14}^{+0.08}$ & $-0.43_{-0.16}^{+0.11}$ & -4.16$\pm$0.11 & 139.23$\pm$1.85$^{2}$ \\ 
J04393364+2359212 & J04393364+2359212 & -9  & N & ... & 3.25$\pm$0.09 & 11 & M5 & 12 & $-0.92_{-0.29}^{+0.19}$ & $-1.03_{-0.49}^{+0.35}$ & -3.35$\pm$0.12 & 126.69$\pm$1.64 \\ 
J04394488+2601527 & ITG 15 & -9  & N & ... & 3.61$\pm$0.10 & 11 & M5.0 & 14 & $-0.92_{-0.29}^{+0.19}$ & $-1.03_{-0.49}^{+0.35}$ & -3.24$\pm$0.12 & 136.37$\pm$2.08 \\ 
J04394748+2601407 & CFHT 4 & 0  & N & ... & 2.00$\pm$0.50 & 8 & M7 & 19 & $-1.41_{-0.18}^{+0.20}$ & $-1.83_{-0.25}^{+0.31}$ & -3.24$\pm$0.14 & 146.82$\pm$5.16 \\ 
J04400067+2358211 & J04400067+2358211 & 0  & Y & ... & 3.13$\pm$0.08 & 11 & M6 & 12 & $-1.22_{-0.20}^{+0.29}$ & $-1.52_{-0.31}^{+0.49}$ & -3.29$\pm$0.14 & 120.15$\pm$2.31 \\ 
J04400800+2605253 & IRAS 04370+2559  & 0  & Y & ... & 51.70$\pm$0.80 & 8 & M2.9 & 36 & $-0.52_{-0.19}^{+0.09}$ & $-0.48_{-0.19}^{+0.11}$ & -2.22$\pm$0.10 & 136.66$\pm$8.50 \\ 
J04403979+2519061 & J04403979+2519061 & -1  & N & ... & $<$0.43 & 10 & M5.25/7 & 13/8 & $-0.68_{-0.27}^{+0.21}$ & $-1.06_{-0.39}^{+0.38}$ & $<$-4.16 & 135.73$\pm$0.05$^{2}$ \\ 
J04404950+2551191 & JH 223 A & 1  & Y &  2.100$^{6}$ & 1.70$\pm$0.14 & 7 & M2.8 & 14 & $-0.51_{-0.18}^{+0.09}$ & $-0.47_{-0.19}^{+0.11}$ & -3.69$\pm$0.09 & 139.39$\pm$1.08 \\ 
J04404950+2551191 & JH 223 B & 2  & Y &  2.100$^{6}$ & 0.76$\pm$0.13 & 7 & M6 & 38 & $-1.22_{-0.20}^{+0.29}$ & $-1.52_{-0.31}^{+0.49}$ & -3.78$\pm$0.15 & 139.39$\pm$1.08$^{1}$ \\ 
J04410826+2556074 & ITG 33A & 0  & Y & ... & 4.10$\pm$0.23 & 7 & M3 & 39 & $-0.54_{-0.19}^{+0.10}$ & $-0.50_{-0.19}^{+0.12}$ & -3.30$\pm$0.09 & 140.48$\pm$4.19 \\ 
J04411078+2555116 & ITG 34 & 0  & Y & ... & 0.70$\pm$0.12 & 7 & M5.5 & 13 & $-1.05_{-0.27}^{+0.23}$ & $-1.25_{-0.45}^{+0.39}$ & -3.78$\pm$0.14 & 156.54$\pm$6.29 \\ 
J04411681+2840000 & CoKu Tau/4 AB & -1  & N & ... & 3.40$\pm$1.90 & 8 & M1.1/2.5 & 8 & $-0.06_{-0.09}^{+0.10}$ & $-0.04_{-0.14}^{+0.13}$ & -3.50$\pm$0.22 & 138.80$\pm$18.70$^{3}$ \\ 
J04412464+2543530 & ITG 40 & -9  & N & ... & 1.00$\pm$0.12 & 7 & M3.5 & 24 & $-0.62_{-0.20}^{+0.14}$ & $-0.59_{-0.26}^{+0.16}$ & -3.90$\pm$0.10 & 138.77$\pm$1.19$^{2}$ \\ 
J04413882+2556267 & IRAS 04385+2550  & 0  & Y & ... & 25.50$\pm$2.20 & 8 & M0 & 40 & $-0.25_{-0.10}^{+0.04}$ & $-0.11_{-0.14}^{+0.06}$ & -2.61$\pm$0.09 & 139.22$\pm$1.28$^{2}$ \\ 
J04414489+2301513 &  J04414489+2301513 & 0  & N & ... & $<$0.38 & 7 & M8.5 & 17 & $-1.67_{-0.13}^{+0.17}$ & $-2.20_{-0.19}^{+0.25}$ & $<$-4.03 & 120.45$\pm$5.54 \\ 
J04414825+2534304 & J04414825+2534304 & 0  & N & ... & 1.34$\pm$0.09 & 11 & M7.75 & 13 & $-1.55_{-0.15}^{+0.17}$ & $-2.02_{-0.24}^{+0.24}$ & -3.43$\pm$0.10 & 135.81$\pm$3.75 \\ 
J04420777+2523118 & V955 Tau A & 1  & Y &  0.320$^{1}$ & 2.20$\pm$0.17 & 7 & K7 & 16 & $-0.19_{-0.01}^{+0.04}$ & $-0.02_{-0.01}^{+0.07}$ & -3.68$\pm$0.08 & 142.25$\pm$2.38$^{2}$ \\ 
J04420777+2523118 & V955 Tau B & 2  & Y &  0.320$^{1}$ & 0.86$\pm$0.14 & 7 & M2.5 & 16 & $-0.48_{-0.14}^{+0.08}$ & $-0.43_{-0.16}^{+0.11}$ & -3.98$\pm$0.10 & 142.25$\pm$2.38$^{2}$ \\ 
J04422101+2520343 & CIDA 7 & 0  & Y & ... & 14.50$\pm$7.20 & 8 & M5.1 & 14 & $-0.94_{-0.28}^{+0.20}$ & $-1.07_{-0.45}^{+0.37}$ & -2.63$\pm$0.20 & 135.70$\pm$2.28 \\ 
J04423769+2515374 & DP Tau A & 1  & Y &  0.110$^{5}$ & 2.10$\pm$0.23 & 7 & M0.8 & 14 & $-0.33_{-0.10}^{+0.09}$ & $-0.22_{-0.14}^{+0.12}$ & -3.65$\pm$0.09 & 142.08$\pm$1.87$^{2}$ \\ 
J04423769+2515374 & DP Tau B & 2  & Y &  0.110$^{5}$ & 1.50$\pm$0.21 & 7 & M2 & 8 & $-0.44_{-0.10}^{+0.09}$ & $-0.38_{-0.12}^{+0.13}$ & -3.75$\pm$0.10 & 142.08$\pm$1.87$^{2}$ \\ 
J04430309+2520187 & GO Tau & 0  & Y & ... & 53.20$\pm$2.80 & 8 & M2.3 & 14 & $-0.46_{-0.12}^{+0.09}$ & $-0.41_{-0.14}^{+0.12}$ & -2.18$\pm$0.09 & 143.98$\pm$1.00 \\ 
J04432023+2940060 &  J04432023+2940060 & 0  & Y & ... & $<$0.38 & 7 & M5.5 & 14 & $-1.05_{-0.27}^{+0.23}$ & $-1.25_{-0.45}^{+0.39}$ & $<$-3.98 & 170.46$\pm$3.65 \\ 
J04442713+2512164 & IRAS 04414+2506 & 0  & N & ... & 4.90$\pm$0.40 & 8 & M7.25 & 13 & $-1.45_{-0.18}^{+0.20}$ & $-1.89_{-0.26}^{+0.32}$ & -2.87$\pm$0.11 & 140.53$\pm$2.71 \\ 
J04455134+1555367 & IRAS 04429+1550 & -9  & N & ... & 51.00$\pm$2.00 & 8 & M2.5 & 35 & $-0.48_{-0.14}^{+0.08}$ & $-0.43_{-0.16}^{+0.11}$ & -2.18$\pm$0.09 & 147.38$\pm$1.25 \\ 
J04465305+1700001 & DQ Tau AB  & -1  & N & ... & 69.30$\pm$4.70 & 8 & M0/0 & 41 & $0.05_{-0.10}^{+0.04}$ & $0.19_{-0.14}^{+0.06}$ & -1.95$\pm$0.09 & 196.36$\pm$2.01 \\ 
J04465897+1702381 & Haro 6-37 A & 1  & N & ... & 2.10$\pm$0.34 & 7 & K8.0 & 14 & $-0.20_{-0.01}^{+0.01}$ & $-0.03_{-0.01}^{+0.01}$ & -3.42$\pm$0.10 & 195.65$\pm$0.86$^{2}$ \\ 
J04465897+1702381 & Haro 6-37 B & 2  & N & ... & $<$0.96 & 7 & M0.9 & 14 & $-0.34_{-0.09}^{+0.10}$ & $-0.23_{-0.14}^{+0.13}$ & $<$-3.71 & 195.65$\pm$0.86$^{2}$ \\ 
J04465897+1702381 & Haro 6-37 C & 3  & N & ... & 38.50$\pm$1.95 & 7 & M1 & 20 & $-0.35_{-0.09}^{+0.10}$ & $-0.25_{-0.13}^{+0.14}$ & -2.10$\pm$0.08 & 195.65$\pm$0.86$^{2}$ \\ 
J04470620+1658428 & DR Tau  & 0  & Y & ... & 115.20$\pm$6.90 & 8 & K6 & 14 & $-0.15_{-0.04}^{+0.05}$ & $0.05_{-0.07}^{+0.08}$ & -1.70$\pm$0.08 & 194.60$\pm$2.46 \\ 
J04474859+2925112 & DS Tau  & 0  & Y & ... & 16.50$\pm$1.80 & 8 & M0.4 & 14 & $-0.28_{-0.10}^{+0.07}$ & $-0.16_{-0.15}^{+0.09}$ & -2.67$\pm$0.09 & 158.35$\pm$1.12 \\ 
J04514737+3047134 & UY Aur A & 1  & Y &  0.880$^{1}$ & 20.80$\pm$0.23 & 9 & M0 & 16 & $-0.25_{-0.10}^{+0.04}$ & $-0.11_{-0.14}^{+0.06}$ & -2.60$\pm$0.08 & 154.93$\pm$1.43 \\ 
J04514737+3047134 & UY Aur B & 2  & Y &  0.880$^{1}$ & 7.87$\pm$0.25 & 9 & M2.5 & 16 & $-0.48_{-0.14}^{+0.08}$ & $-0.43_{-0.16}^{+0.11}$ & -2.95$\pm$0.09 & 154.93$\pm$1.43$^{1}$ \\ 
J04542368+1709534 & St 34 Aab & -1  & N & ... & $<$4.20 & 8 & M3/3 & 32 & $-0.24_{-0.19}^{+0.10}$ & $-0.19_{-0.19}^{+0.12}$ & $<$-3.35 & 142.16$\pm$1.15 \\ 
J04542368+1709534 & St 34 B & 2  & N & ... & $<$4.20 & 8 & M5.5 & 42 & $-1.05_{-0.27}^{+0.23}$ & $-1.25_{-0.45}^{+0.39}$ & $<$-3.09 & 142.16$\pm$1.15$^{1}$ \\ 
J04554535+3019389 &  J04554535+3019389 & 0  & Y & ... & $<$0.39 & 7 & M4.7 & 14 & $-0.86_{-0.36}^{+0.20}$ & $-0.92_{-0.60}^{+0.29}$ & $<$-4.13 & 154.12$\pm$2.33 \\ 
J04554801+3028050 &  J04554801+3028050 & 2  & N & ... & $<$0.35 & 7 & M5.6 & 13 & $-1.06_{-0.29}^{+0.22}$ & $-1.25_{-0.49}^{+0.36}$ & $<$-4.08 & 156.19$\pm$3.73 \\ 
J04554969+3019400 &  J04554969+3019400 & -9  & N & ... & $<$0.35 & 7 & M6 & 13 & $-1.22_{-0.20}^{+0.29}$ & $-1.52_{-0.31}^{+0.49}$ & $<$-4.02 & 156.05$\pm$4.63 \\ 
J04555605+3036209 & XEST 26-062 & -9  & N & ... & 0.65$\pm$0.14 & 11 & M4.0 & 14 & $-0.73_{-0.19}^{+0.19}$ & $-0.68_{-0.35}^{+0.19}$ & -3.96$\pm$0.12 & 156.58$\pm$0.86$^{2}$ \\ 
J04555938+3034015 & SU Aur  & 0  & Y & ... & 27.40$\pm$2.50 & 8 & G4 & 14 & $0.49_{-0.02}^{+0.01}$ & $1.32_{-0.09}^{+0.10}$ & -2.83$\pm$0.09 & 157.68$\pm$1.48 \\ 
J04560118+3026348 & XEST 26-071 & -9  & N & ... & $<$0.35 & 7 & M3.1 & 14 & $-0.55_{-0.20}^{+0.10}$ & $-0.51_{-0.19}^{+0.12}$ & $<$-4.27 & 156.58$\pm$0.86$^{2}$ \\ 
J05030659+2523197 & V836 Tau  & 0  & Y & ... & 29.10$\pm$2.40 & 8 & M0.8 & 14 & $-0.33_{-0.10}^{+0.09}$ & $-0.22_{-0.14}^{+0.12}$ & -2.36$\pm$0.09 & 168.76$\pm$1.24 \\ 
J05044139+2509544 & CIDA 8 & 0  & Y & ... & 7.70$\pm$1.60 & 8 & M3.7 & 14 & $-0.67_{-0.20}^{+0.17}$ & $-0.63_{-0.29}^{+0.18}$ & -2.82$\pm$0.12 & 169.99$\pm$2.38 \\ 
J05052286+2531312 & CIDA 9 A & 1  & Y &  2.300$^{6}$ & 33.80$\pm$1.69 & 7 & M1.8 & 14 & $-0.42_{-0.09}^{+0.10}$ & $-0.36_{-0.11}^{+0.14}$ & -2.24$\pm$0.08 & 171.08$\pm$2.62 \\ 
J05052286+2531312 & CIDA 9 B & 2  & Y &  2.300$^{6}$ & $<$0.42 & 7 & M4.6 & 14 & $-0.84_{-0.22}^{+0.20}$ & $-0.89_{-0.36}^{+0.28}$ & $<$-4.02 & 171.08$\pm$2.62$^{1}$ \\ 
J05062332+2432199 & CIDA 11 & -1  & N & ... & $<$0.36 & 7 & M4/4.5 & 43/8 & $-0.43_{-0.19}^{+0.19}$ & $-0.46_{-0.36}^{+0.22}$ & $<$-4.37 & 138.80$\pm$18.70$^{3}$ \\ 
J05074953+3024050 & RW Aur A  & 1  & Y &  1.420$^{1}$ & 27.20$\pm$2.20 & 8 & K0 & 14 & $0.20_{-0.04}^{+0.08}$ & $0.64_{-0.07}^{+0.14}$ & -2.77$\pm$0.13 & 138.80$\pm$18.70$^{3}$ \\ 
J05074953+3024050 & RW Aur B  & 2  & Y &  1.420$^{1}$ & 4.40$\pm$0.80 & 8 & K6.5 & 14 & $-0.18_{-0.02}^{+0.07}$ & $0.00_{-0.03}^{+0.10}$ & -3.40$\pm$0.14 & 138.80$\pm$18.70$^{3}$ \\ 
J05075496+2500156 & CIDA 12 & 0  & Y & ... & 1.16$\pm$0.09 & 11 & M3.7 & 14 & $-0.67_{-0.20}^{+0.17}$ & $-0.63_{-0.29}^{+0.18}$ & -3.68$\pm$0.10 & 164.15$\pm$2.39 \\ 

\enddata
\tablenotetext{a}{Role values: 0 = verified single, -9 = unknown, 1 = primary, 2 = secondary, 3 = tertiary, 4 = quaternary, -1 = SB primary, -2 = SB secondary }
\tablenotetext{b}{Separation, flux, and spectral type references: 1) \citet{whi01}; 2) \citet{kh12}; 3) \citet{cor06}; 4) \citet{kra09}; 5) \citet{kra11}; 6) \citet{kra07}; 7) This work; 8) \citet{and13}; 9) Paper 1; 10) \citet{har15}; 11) \citet{wd18}; 12) \citet{fur11}; 13) \citet{luh04}; 14) \citet{HH2014}; 15) \citet{Bri98}; 16) \citet{Har03}; 17) \citet{Luhman2006a}; 18) \citet{Luhman2003}; 19) \citet{Bri02}; 20) \citet{har94}; 22) \citet{Canty2013}; 23) \citet{Cieza2012}; 24) \citet{Luhman2006b}; 25) \citet{Luhman2009b}; 26) \citet{Beck2007}; 27) \citet{Espaillat2010}; 28) \citet{duc99}; 29) \citet{Monin1998}; 30) \citet{White2004}; 31) \citet{White1999}; 32) \citet{Prato2002}; 33) \citet{Schaefer2012}; 34) \citet{Walter2003}; 35) \citet{Luhman2009a}; 36) \citet{luh10}; 37) \citet{Slesnick2006}; 38) \citet{Esplin2018}; 39) \citet{Martin2000}; 40) \citet{Kenyon1998}; 41) \citet{mat97}; 42) \citet{Dahm2011}; 43) \citet{Kenyon1995} } 
\tablenotetext{c}{Distance flag: none: direct Gaia match; 1: used brightest Gaia source within 8\arcsec\ for companions in multiple systems 2: No Gaia distance or high RUWE, used weighted mean of Gaia distance stars in our sample within 30’;  3: no Gaia detections within 30\arcmin in our sample, used median Class II sample Gaia distance and standard deviation of 138.8 $\pm$ 18.7 pc}
\end{deluxetable}
\end{longrotatetable}

\end{document}